\documentclass[11pt]{article}

\usepackage{cite}
\usepackage{hyperref}

\usepackage{lineno}

\usepackage{amsmath}
\usepackage{amssymb,amsfonts,amsthm}
\usepackage{epsfig,graphicx}
\usepackage{url} 
\usepackage{bm} 
\usepackage{slashed}
\usepackage{cite}

\usepackage{color}
\usepackage{pstricks}

\usepackage{fancyhdr}
\usepackage{textcomp}
\newcommand{\unit}{\leavevmode\hbox{\small1\kern-3.6pt\normalsize1}}

\def \GeV{{\mathrm{GeV}}}

\def\CM {{\cal M}}

\def\CM {{\cal M}}

\def\tanb{\tan\beta}

\newcommand{\sinbsq}{\sin^2\beta}
\newcommand{\props}[1]{s-\MHsq{#1}+i\MH{#1}\GamH{#1}}

\def\mz{M_Z}
\def\mw{M_W}

\def\snr{{\tilde N}}

\def\snmassr{{m_{\tilde N_1}}}

\def\chisnsn{C_{\higgsi\snr_1\snr_1}}
\def\chjsnsn{C_{\higgsj\snr_1\snr_1}}

\newcommand{\chcc}[1]{C_{\chiggsp\chiggsm\higgs{#1}}}

\def\cccsnsn{C_{\chiggsp\chiggsm\snr_1\snr_1}}

\def\al{{A_\lambda}}
\def\ak{{A_\kappa}}

\def\ln{{\lambda_N}}
\def\aln{{A_{\lambda_N}}}
\def\mn{{m_{\tilde{N}}}}

\def\hsm{h_{SM}^0}

\def\higgsi{{H_i^0}}
\def\higgsj{{H_j^0}}

\def\higgsl{{H_1^0}}

\def\hcompiu{S_{H_i^0}^2}

\def\hcompju{S_{H_j^0}^2}

\def\phiggsl{{A_1^0}}

\def\phmassl{{m_{A_1^0}}}

\def\chiggsp{{H^+}}
\def\chiggsm{{H^-}}
\def\chiggspm{{H^\pm}}

\def\ln{{\lambda_N}}
\def\aln{{A_{\lambda_N}}}
\def\mn{{m_{\tilde{N}}}}

\newcommand{\alpem}{\alpha_{em}}
\newcommand{\cprops}[1]{s-\MHsq{#1}-i\MH{#1}\GamH{#1}}

\newcommand{\taufrac}[1]{\frac{4m_{#1}^2}{s}}

\newcommand{\MH}[1]{m_{H^0_{#1}}}
\newcommand{\MHsq}[1]{\MH{#1}^2}
\newcommand{\GamH}[1]{\Gamma_{H^0_{#1}}}

\newcommand{\hs}[1]{H^0_{#1}}
\newcommand{\ch}[1]{C_{\hs{#1}}}

\newcommand{\chg}{\tilde \chi^\pm}

\newcommand{\mcha}[1]{m_{\tilde \chi^\pm_{#1}}}
\newcommand{\ccchch}[3]{C_{H^0_{#1}\tilde{\chi}^{+}_{#2}\tilde{\chi}^{-}_{#3}}}

\newcommand{\Qab}[3]{Q_{#1#2#3}}

\newcommand{\st}{\tilde{f}}

\newcommand{\chstst}[2]{C_{H^0_{#1}\st_{#2}\st_{#2}}}

\newcommand{\higgs}[1]{H_{#1}^0}

\newcommand{\tauh}{\tau_{H^\pm}}
\newcommand{\taut}{\tau_{t}}
\newcommand{\tauw}{\tau_{W}}
\newcommand{\taufa}{\tau_{f_a}}
\newcommand{\taufb}{\tau_{f_b}}
\newcommand{\tauxa}{\tau_{\tilde\chi^{\pm}_a}}
\newcommand{\tauxb}{\tau_{\tilde\chi^{\pm}_b}}

\def\neutl{\widetilde \chi^0_1}
\def\neutlmass{m_{\widetilde \chi^0_1}}

\def \GeV{{\mathrm{GeV}}}

\def\sigmav{\langle\sigma v\rangle_0}
\newcommand{\bsg}{b\to s\gamma}
\newcommand{\bmumu}{B_S\to\mu^+\mu^-}
\newcommand{\btaunu}{B^+ \to \tau^+ \nu_\tau}

\allowdisplaybreaks

\def\sigmavgg{\xi^2\langle\sigma v\rangle_{\gamma\gamma} }
\def\sineut{\xi\sigma^{SI}_{\tilde\chi_1-p}}
\def\sdpneut{\xi\sigma^{SD}_{\tilde\chi_1-p}}
\def\sdnneut{\xi\sigma^{SD}_{\tilde\chi_1-n}}

\begin{document}

\thispagestyle{empty}
\begin{flushright}

 {\small IPPP/15/50; 
 DCTP/15/100\\
 IFT-UAM/CSIC-15-82; 
 FTUAM-15-23}\\

  \vspace*{2.mm}{July 31, 2015}
\end{flushright}

\begin{center}
  {\bf {\LARGE Enhanced lines and box-shaped features in the gamma-ray spectrum from annihilating dark matter in the NMSSM}}

\renewcommand*{\thefootnote}{\fnsymbol{footnote}}
\setcounter{footnote}{3}

  \vspace{0.5cm}
  {\large
    D.~G.~Cerde\~no $^{a,}$\footnote{davidg.cerdeno@gmail.com},
    \setcounter{footnote}{0}
    M. Peir\'o $^{b,c,}$\footnote{miguel.peiro@uam.es} and
    \setcounter{footnote}{6}
    S. Robles $^{b,c,}$\footnote{sandra.robles@uam.es} 
  }
  \\[0.2cm] 

  {\footnotesize{
	$^a$ Institute for Particle Physics Phenomenology, Department of Physics\\
	Durham University, Durham DH1 3LE, United Kingdom\\
       $^b$ Instituto de F\'{\i}sica Te\'{o}rica UAM/CSIC, Universidad Aut\'{o}noma de Madrid, 28049 Madrid, Spain\\
       $^c$ Departamento de F\'{\i}sica Te\'{o}rica,
       Universidad Aut\'{o}noma de Madrid, 28049
       Madrid, Spain\\ 
        }
    }

\vspace*{0.7cm}

  \begin{abstract}
    \noindent
We study spectral features in the gamma-ray emission from dark matter (DM) annihilation in the Next-to-Minimal Supersymmetric Standard Model (NMSSM), with either neutralino or right-handed (RH) sneutrino DM. 
We perform a series of scans over the NMSSM parameter space, compute the DM annihilation cross section into two photons and the contribution of box-shaped features, and compare them with the limits derived from the Fermi-LAT search for gamma-ray lines using the latest Pass 8 data. 
We implement the LHC bounds on the Higgs sector and on the masses of supersymmetric particles as well as the constraints on low-energy observables. 
We also consider the recent upper limits from the Fermi-LAT satellite on the continuum gamma-ray emission from dwarf spheroidal galaxies (dSphs).
We show that in the case of the RH sneutrino the constraint on gamma-ray spectral features can be more stringent than the dSph bounds.
This is due to the Breit-Wigner enhancement near the ubiquitous resonances with a CP even Higgs and the contribution of scalar and pseudoscalar Higgs final states to box-shaped features.
By contrast, for neutralino DM, the di-photon final state is only enhanced in the resonance with a $Z$ boson and box-shaped features are even more suppressed.
Therefore, the observation of spectral features could constitute a discriminating factor between both models.
In addition, we compare our results with direct DM searches, including the SuperCDMS and LUX limits on the elastic DM-nucleus scattering cross section and show that some of these scenarios would be accessible to next generation experiments.
Thus, our findings strengthen the idea of complementarity among distinct DM search strategies.
  \end{abstract}
\end{center}

\newpage


\section{Introduction}
\setcounter{footnote}{0}
\renewcommand*{\thefootnote}{\arabic{footnote}}

The existence of a vast amount of dark matter (DM) in the Universe is supported by many different and independent observations.
The latest measurements suggest that approximately a 27\% of the Universe energy density is in form of a new type of non-baryonic cold DM~\cite{Ade:2013zuv}. 
Given that the Standard Model (SM) of particle physics does not contain any viable candidate to account for it, DM can be regarded as one of the clearest hints of new physics. 
Among the many diverse proposals within particle physics, Weakly Interacting Massive Particles (WIMPs), provide a very well-motivated paradigm, since these particles can be thermally produced in the early Universe in a sufficient amount to reproduce the observed relic abundance.

The hunt for DM is an extremely active field. These exotic particles could be observed through their collisions with nuclei inside direct detection experiments, a challenging technique that involves the use of very sensitive underground detectors. 
They could also be produced at colliders, such as the LHC, where they are expected to leave an imbalance in the transverse energy as they escape the detectors. 
Finally, the annihilation (or decay) of DM particles in the Galactic halo as well as in other astrophysical objects, and the subsequent emission of SM particles (gamma-rays, antiparticles or neutrinos) is being searched for by a variety of techniques, collectively referred to as indirect detection. The challenge in the latter strategy is the discrimination between a potential signal and the astrophysical background, which requires a good modelling of all the conventional physical processes that can produce cosmic rays and photons.

Although various potential hints of DM detection have been reported in the last decades, both from direct as well as indirect detection experiments, none of these have been conclusively confirmed yet.
To their credit, the DAMA/LIBRA annual modulation \cite{Bernabei:2010mq}; the $511$~keV excess observed by the SPI experiment \cite{Jean:2003ci}; the PAMELA and AMS rise in the positron fraction \cite{Adriani:2008zr,Aguilar:2013qda}; a potential X-ray line emission near 3.5 keV observed in our Galaxy and certain galaxy clusters \cite{Bulbul:2014sua,Boyarsky:2014jta}; and the apparent excess in Fermi-LAT data from gamma-rays in the Galactic Centre (GC) \cite{Vitale:2009hr,Hooper:2010mq,Morselli:2010ty,Hooper:2011ti,Abazajian:2012pn,Daylan:2014rsa,Gordon:2013vta,Abazajian:2014fta,Zhou:2014lva,Calore:2014xka} have triggered an enthusiastic examination of DM models beyond the ``vanilla" scenario during the past decade. 
Among these potential signatures, the Fermi-LAT low-energy excess in the Galactic Centre Emission (GCE) \cite{TheFermi-LAT:2015kwa} stands out, as it could be explained in terms of WIMP DM, with a mass in the $6-200$~GeV range and an annihilation cross section compatible with that of thermal relics\footnote{Although an astrophysical explanation in terms of millisecond pulsars, or leptonic cosmic ray outbursts seems also plausible \cite{Cholis:2015dea}.}.
In any case, the verification of this and other DM hints would need an independent complementary measurement, which would allow us to gain further insight into the DM particle nature.

In this regard, the observation of spectral features in the gamma-ray spectrum 
would constitute an excellent discrimination tool, as these features are linked to characteristic properties of DM models. 
For example, the energy of monochromatic photons from either the annihilation or decay of DM particles could provide a measurement of the DM mass.
Similarly, box-shaped features and internal bremsstrahlung processes can be associated to specific annihilation channels.
Furthermore, since in general, astrophysical processes predict continuous $\gamma$-ray spectra at the GeV energy scale, this observation would be a smoking gun of DM detection.

The Fermi Large Area Telescope (Fermi-LAT) is currently performing an all-sky survey in $\gamma$-rays with unprecedented sensitivity in the GeV range.
Specially interesting for DM searches is the GC region, where a high density of DM particles is expected and 
therefore higher gamma-ray fluxes from their annihilation or decay are predicted, as well as nearby dwarf spheroidal galaxies (dSphs), which are DM dominated objects. 
No DM signal has been obtained from the latter, leading to very stringent upper bounds on the current DM annihilating cross section in the Galactic halo, $\sigmav$, which rules out the ``thermal" value ($\sigmav\approx3\times10^{-26}$~cm$^3$/s) for WIMP masses below $\sim$~100~GeV (the precise value depends on the main annihilation channel). 
These constraints are in mild tension with the DM interpretation of the GCE that we have briefly mentioned above \cite{Calore:2014nla,Ackermann:2015zua}.

Regarding the search for features in the gamma-ray spectrum, no significant global excess of a spectral line has been found yet. 
A recent analysis of 5.8 years of data (Pass 8) \cite{Ackermann:2015lka} has explored the energy range from 200~MeV to 500 GeV for five different regions of interest (ROIs), optimised to maximise the signal to noise ratio. A previous study had extended the search to energies as low as 100~MeV \cite{Albert:2014hwa} using the Pass 7 data set.
Upper constraints on the annihilation cross section of DM particles into monochromatic photons have been derived in the range $\langle\sigma v\rangle_{\gamma\gamma} \sim 10^{-29}-10^{-27}$ cm$^3/$s, depending on the WIMP mass and the DM density profile assumed. 
These limits begin to explore the expected values for a loop-suppressed WIMP annihilation cross section and thus constrain some regions of the parameter space of DM models.

From the perspective of particle models for DM, the annihilation of WIMPs into a pair of photons is a loop-suppressed process,  hence it is
expected to be three or more orders of magnitude below the thermal cross section \cite{Bergstrom:1988fp,Bergstrom:1997fh,Bern:1997ng,Ullio:1997ke,
Bergstrom:1997fj,
Bergstrom:2004nr,Boudjema:2005hb,Ferrer:2006hy,Gustafsson:2007pc,Bertone:2009cb,Bringmann:2011ye,Chalons:2011ia,Bringmann:2012ez}.   
However, there are known mechanisms by which the line emission can be enhanced, which were explored in order to account for a previous apparent excess at 130~GeV  \cite{Bringmann:2012vr,Weniger:2012tx,Tempel:2012ey,Boyarsky:2012ca,Kang:2012bq,Chatterjee:2014bva} that is no longer statistically significant. For instance, resonant annihilation, together with annihilation just above the production threshold of the particles involved in the loop \cite{Jackson:2013pjq}  can significantly increase the rate of this process.
Furthermore, gamma rays can also be produced in the decay in-flight of certain annihilation final states (such as scalar and pseudoscalar Higgs bosons), giving rise to characteristic box-shaped features in the gamma-ray spectrum.
Finally, gamma-ray features from internal bremsstrahlung \cite{Bergstrom:1989jr,Flores:1989ru}, linked to processes in which charged particles mediate DM annihilation, can also lead to observable signatures. 
The theoretical predictions for these processes are extremely model dependent and, as such, could be used in order to distinguish between different DM scenarios.

In this article, we work within the context of the
Next-to-Minimal Supersymmetric Standard Model (NMSSM).
The NMSSM is a well-motivated extension of the Minimal Supersymmetric Standard Model in which an extra singlet field, $S$, is introduced, which mixes with the Higgs SU$(2)$ doublet and whose vacuum expectation value after electroweak symmetry breaking generates an effective EW-scale $\mu$ parameter \cite{Kim:1983dt} (see, e.g., Ref.\,\cite{Ellwanger:2009dp} for a review).
Among its many virtues, the NMSSM displays a very interesting phenomenology, mainly due to its enlarged Higgs sector. The mixing of the Higgs doublet with the new singlet component opens the door to very light scalar and pseudoscalar Higgs bosons with interesting prospects for collider searches.
Moreover, in the NMSSM the mass of the Higgs boson also receives new tree-level contributions from the new terms in the superpotential \cite{Cvetic:1997ky,Barger:2006dh}, which allow to increase the Higgs mass and reduce the amount of fine-tuning of the model \cite{BasteroGil:2000bw,Ellwanger:2014dfa,Kaminska:2014wia}. 
The profound changes in the neutralino and Higgs sectors also have interesting implications for neutralino DM and new contributions to both direct and indirect detection. 
The NMSSM can be enlarged with an extra singlet superfield that incorporates right-handed neutrinos (and sneutrinos) \cite{Kitano:1999qb,Deppisch:2008bp} in order to accommodate a see-saw mechanism that explains the smallness of neutrino masses. The right-handed (RH) sneutrino in the resulting construction is a viable DM candidate \cite{Cerdeno:2008ep} with interesting phenomenological properties.

In this paper, we compute the annihilation cross section into two photons and the contribution from box-shaped features for the lightest neutralino and the RH sneutrino in the aforementioned supersymmetric constructions, focusing on the low-mass region below 
$200$~GeV. 
We compare the results with the current upper bounds derived from the search for features in the gamma-ray spectrum studied by the Fermi-LAT collaboration. 
We investigate the effect of these constraints and compare them with Fermi-LAT limits on the continuum emission of gamma-rays from dSphs, as well as direct detection bounds on the elastic scattering cross section. 
It is important to emphasize that we do not make any attempt to fit the GCE in this paper. 
We observe that for certain regions of the parameter space with RH sneutrino DM, 
the search for these spectral features can be more constraining than the existing bounds from dSphs, and even than direct detection limits.  
This rarely happens for the case of the neutralino, unless the narrow resonant annihilation condition with the $Z$ boson is fulfilled.
These results emphasize the relevance of this technique in order to discriminate among DM candidates.

This article is organised as follows. 
In Section\,\ref{sec:neutralino} we investigate the NMSSM with neutralino DM and in Section\,\ref{sec:sneutrino} we address the case of the RH sneutrino in an extended NMSSM.
Finally, the conclusions are presented in Section \ref{sec:conclusions}.


\section{Spectral features from neutralino annihilation in the NMSSM}
\label{sec:neutralino}

The modifications in the neutralino sector due to the inclusion of a singlino component and the presence of new annihilation channels have profound consequences for DM searches, which have been extensively discussed in the literature \cite{Menon:2004wv,Cerdeno:2004xw,Belanger:2005kh,Cerdeno:2007sn,Hugonie:2007vd}. 
For example, it has been shown that the NMSSM can accommodate low-mass neutralino DM~\cite{Gunion:2005rw,Ferrer:2006hy,Aalseth:2008rx,Vasquez:2010ru,Cao:2011re,AlbornozVasquez:2011js,Kozaczuk:2013spa,Cao:2013mqa}. 
Several studies have recently pointed out that these neutralinos can also explain the GCE~\cite{Cheung:2014lqa,Huang:2014cla,Cao:2014efa,Gherghetta:2015ysa,Cao:2015loa} when they are either singlino-Higgsino or bino-Higgsino admixtures~\cite{Cheung:2014lqa}.

The monochromatic gamma-ray spectrum from neutralino annihilation in the context of the NMSSM has been previously studied in Refs.\,\cite{Ferrer:2006hy,Chalons:2011ia,Lee:2012bq,Das:2012ys,Chalons:2012xf}. In general, one expects that the cross section of the process $\tilde{\chi}^0_1\tilde{\chi}^0_1\rightarrow\gamma\gamma$ is loop suppressed with respect to the thermal annihilation cross section. In Ref.~\cite{Ferrer:2006hy}, the authors showed that for neutralinos lighter than 100 GeV in the NMSSM, the cross section for the $\gamma\gamma$ final state is typically of the order of $10^{-31}$~cm$^3$\,s$^{-1}$ for either bino-like, singlino-like or mixed neutralinos. 
This value can increase several orders of magnitude when neutralinos have a resonant annihilation mediated by a very light CP-odd Higgs with a fermion loop, reaching  $\langle\sigma v\rangle_{\gamma\gamma}\approx10^{-27}$~cm$^3$\,s$^{-1}$~\cite{Ferrer:2006hy,Lee:2012bq,Das:2012ys,Chalons:2012xf}. 

In this section, we update these results, incorporating the LHC constraints on the Higgs sector. They set stringent bounds on the Higgs couplings and the invisible branching ratio and therefore also limit the availability of very light scalars and pseudoscalars, needed in order for the light neutralinos to fulfil the relic density constraint \cite{Vasquez:2010ru,Cao:2011re,Vasquez:2012hn}. In addition, we take into account the most recent bounds from direct and indirect DM searches. To this aim, we first compute the monochromatic gamma-ray spectrum for a series of scans over the parameter space and determine the observability of this signal. Then, we compare the results with other direct and indirect search strategies. 

When computing the continuum contribution to the gamma ray spectrum, we must also take into account virtual internal bremsstrahlung (VIB) processes that can take place in $t$-channel exchange of a charged particle. In the case of the neutralino, this can correspond to annihilation channels into lepton pairs (mediated by sleptons), $H^+H^-$ (mediated by charged Higgses), or $W^+W^-$ pairs (mediated by charginos).
The contribution from VIB can exceed that from final state radiation\cite{Bergstrom:1989jr,Bergstrom:2005ss} and in some cases be more easily observable than gamma-ray lines \cite{Bringmann:2007nk}.
As it has been shown in Ref.\,\cite{Bringmann:2012vr}, the effect of VIB is enhanced when the mass of the DM and the mass of the charged mediator are almost degenerate. 
We will later argue that these conditions are not satisfied in our scan.

\subsection{Details of the scan and experimental constraints}
\label{sec:neutscan}

We have carried out a series of scans, designed to explore neutralino masses below 200~GeV. The NMSSM input parameters have been defined at the EW scale according to the ranges in Table\,\ref{tab:scan_nmssm}.
Fixed values are used for the gluino soft mass, $M_3=1500$~GeV, for the trilinear parameters,
$A_{U}=3700$~GeV, $A_{D}=2000$~GeV, and $A_{E}=-1000$~GeV, as well
as for the soft scalar masses of sleptons and squarks,
$m_{\tilde{L}_i}=m_{\tilde{E}_i}=300$ GeV and
$m_{\tilde{Q}_i}=m_{\tilde{U}_i}=m_{\tilde{D}_i}=1500$~GeV, respectively, where
the index $i$ runs over the three families. 
The conservative choice of the squark masses is motivated by the LHC null results from supersymmetry searches. Also note that despite the large trilinear term $A_U$, the instability against charge- and/or color-breaking minima is avoided since the soft mass of the squarks is at the TeV scale \cite{Ellwanger:1999bv}.
A large $A_U$ helps to increase the loop contributions to the CP-even Higgs mass and thus is useful for reproducing the experimental value (this is more relevant in the MSSM than in the NMSSM, where extra tree-level contributions are present).
Small slepton masses are helpful to enlarge the supersymmetric contribution to the muon anomalous magnetic moment (although this constraint is not imposed in our analysis). 
We consider that these scans are sufficient to study the phenomenology of monochromatic gamma-ray emission from the annihilation of low-mass neutralinos\footnote{As we will comment later, we do expect that modifications of some of these input parameters have an influence on other aspects of the neutralino phenomenology, such as the contribution of VIB to the continuum gamma-ray spectrum.}.

The range of variation of our parameters in Table\,\ref{tab:scan_nmssm} is similar to the one used in other scans of low-mass neutralino DM in the NMSSM \cite{Vasquez:2010ru,Cao:2011re,AlbornozVasquez:2011js,Vasquez:2012hn,Kozaczuk:2013spa,Cao:2013mqa}.
Some of these analyses have used MCMC techniques to explore the NMSSM parameter space. In the case of Ref.\,\cite{Vasquez:2010ru} 
the authors considered 11 initial parameters. 
We have focused our scan in the regions where more solutions were found in Ref.\,\cite{Vasquez:2010ru}, but adopting fixed values for the scalar masses.
Also, Ref.\,\cite{Cao:2013mqa} extends the range in $|\ak|$ up to $200$~GeV, but in Ref.\,\cite{Vasquez:2010ru} small values of $|\ak|$ are preferred in order to have low-mass neutralinos.

The masses of the supersymmetric particles and low-energy observables have been computed with {\tt NMSSMTools 4.1.2}~\cite{Ellwanger:2004xm,Ellwanger:2005dv,Ellwanger:2006rn}, whereas the gamma-ray spectrum and the DM relic abundance have been computed with {\tt micrOMEGAs 3.6.9}~\cite{Belanger:2013oya}.
In order to efficiently explore the parameter space, we have linked these codes with {\tt MultiNest 3.9} \cite{Feroz:2007kg,Feroz:2008xx,Feroz:2013hea}, which uses a likelihood function to generate MCMC and to find regions of the 
parameter space that maximise the likelihood.
This function contains uncorrelated Gaussian probability distributions for the DM relic abundance, the SM-like Higgs mass, 
${\rm BR}(B_s\to \mu^+\mu^-)$, and ${\rm BR}(b\to s\gamma)$, centred around the observed values.

It is important to stress that the scan procedure has not been optimised to extract any statistical information on the results and should not be regarded as a complete exploration of the vast multidimensional parameter space. 
In this article, we are only concerned with obtaining a representative subset of viable solutions that covers the low-mass range of the DM parameter space without addressing the probability of each point to explain the whole set of experimental data considered.

\begin{table}
  \begin{center}
    \begin{tabular}{|c|c|c|c|}
      \hline
      Parameter & Scan 1 & Scan 2&Scan 3\\
      \hline
      \hline 
      $M_1$& $[1, 200]$ & $[1 , 40]$ & $[1 , 200]$\\
      $M_2$&  $[200 , 1000]$ & $[200 , 1000]$& $[700 , 1000]$\\
      $\tan\beta$ & $[4 , 20]$ & $[4 , 20]$ & $[2 , 50]$\\
      $\lambda$& $[0.1 , 0.6]$ & $[0.1 , 0.6]$& $[0.001 , 0.1]$\\
      $\kappa$& $[0 , 0.1]$ & $[0 , 0.1]$& $[0.1 ,0.6]$\\
      $A_\lambda$& $[500 , 5000]$ & $[500 , 5000]$& $[500 , 1100]$\\    
      $A_\kappa$& $[-50 , 50]$ & $[-30 , 0]$& $[-50 , 50]$ \\    
      $\mu_{eff}$& $[110 , 250]$ & $[160 , 250]$& $[200 , 400]$\\   
      \hline
          \end{tabular}
    \caption{\small Input NMSSM parameters for the series
      of scans used in the neutralino case. Masses and trilinear parameters are given in GeV. All parameters are defined at the EW scale. 
      }
    \label{tab:scan_nmssm}
  \end{center}
\end{table}

For each point of the parameter space, we have 
computed the full supersymmetric spectrum and imposed the most recent experimental constraints from collider searches.
For the sleptons and charginos, we include the lower limits on their masses from LEP through the {\tt micrOMEGAs 3.6.9} code. For all the squarks, we apply a lower limit on the mass of 1.5~TeV independent on the gluino mass, which is in agreement with the latest ATLAS results~\cite{Aad:2015iea}. Besides, our choice for the gaugino soft mass parameter, $M_3$, ensures that the gluino mass is above the current LHC bound.
Regarding the Higgs sector, we have ensured that one of the CP-even Higgs bosons lies in the mass range $123$~GeV$\leq m_{\higgs{2}}\leq 128$~GeV (that accounts for the experimental and theoretical uncertainty\footnote{This range might be enlarged due to two-loop corrections to the Higgs masses in the NMSSM, which can be especially important for large values of $\lambda$ \cite{Goodsell:2014pla,Staub:2015aea}.})
and has properties very close to those of the SM Higgs. 
In particular, we have imposed an upper bound on its invisible branching fraction, ${\rm BR}(\hsm\rightarrow {\rm inv})<0.27$.
 We also have included the bounds from the rare decays $\bmumu$, $\bsg$ and $\btaunu$.
 For further details on how these bounds are implemented, see Ref.~\cite{Cerdeno:2014cda}.

We have also implemented the most recent constraints from DM direct detection experiments. In particular, we have made use of the latest results of LUX \cite{Akerib:2013tjd} and SuperCDMS \cite{Agnese:2013jaa,Agnese:2014aze}, which are the most constraining experiments in the mass range considered along this work\footnote{ For smaller masses, the CRESST collaboration provides the most stringent bound on the elastic scattering cross section of DM particles up to date~\cite{Angloher:2014myn}.}. We have computed the upper bound for these two experiments in each point of the parameter space of the scanned model using the Yellin's maximum gap method at 90\% C.L.~\cite{Yellin:2002xd} following the procedure sketched in Ref.~\cite{Marcos:2015dza}. With the data provided in that work for different DM haloes, the spin-independent and spin-dependent contributions are tested simultaneously using the expected number of events in the corresponding detector, thus deriving consistent limits.
For consistency with the analysis of gamma-ray lines of Ref.~\cite{Ackermann:2015lka}, the direct detection bounds have been extracted assuming the same local DM density, $0.4$~GeV$/$cm$^3$, and the same choices of DM halo density profiles\footnote{Note that the profiles (both Einasto and NFW) considered in Ref.~\cite{Ackermann:2015lka} are slightly different from those of Ref.~\cite{Fornasa:2013iaa}, used to extract the speed distributions in Ref.~\cite{Marcos:2015dza}. However, we have checked that these differences only lead to small deviations in the expected number of events. }.

We have set an upper bound on the DM relic abundance, $\Omega h^2<0.13$,
consistent with the latest Planck results~\cite{Ade:2013zuv}. 
We have also considered the possibility that supersymmetric DM only contribute to a fraction of the total relic density, and set for concreteness a lower bound on the relic abundance, $0.001<\Omega h^2$. 
The fractional density, $\xi=\min[1,\Omega h^2/0.11]$, has been introduced to account for the reduction in the rates for direct and indirect searches (assuming that the DM candidate is present in the DM halo in the same proportion as in the Universe)\footnote{This assumption is valid if another cold DM candidate, e.g., the axion, accounts for the missing  relic density. Changes in the thermal history of the Universe or the presence of DM candidates with different interaction properties, such as self-interacting DM, might modify this hypothesis. }.

Finally, regarding indirect detection, we have also incorporated the Fermi-LAT bounds on dSphs, we remind the reader that we do not attempt to fit the GCE. 
The Fermi-LAT collaboration has performed an analysis of the gamma-ray emission from 25 dSphs using six years of data \cite{Ackermann:2015zua}. 
The absence of a signal can be interpreted as constraints on the annihilation cross section of DM particles. 
It is customary to assume annihilation into pure SM channels in the calculation of these bounds.
Nevertheless, the occurrence of non standard annihilation final states in our DM scenarios prevents us from using these results directly. 
Instead, we have extracted independent 95\% CL upper bounds on $\xi^2\sigmav$ for the six more constraining dwarf galaxies, i.e. the ones with the largest $J$-factors \cite{Ackermann:2013yva} (Coma Berenices, Draco, Segue I, Ursa Major II, Ursa Minor and Willman I). 
We have derived the upper limits from the gamma-ray flux predicted by the specific DM model for each dSph, using the delta-log-likelihood method, the mean values of the $J$-factors and the bin-by-bin likelihood functions provided in Ref.~\cite{Ackermann:2013yva}, without taking into account the uncertainties on the kinematically measured $J$-factors.
Then, we have applied the most restrictive of these limits to our data.  
We have checked that this procedure leads to slightly less stringent bounds than the combined limit from the Fermi-LAT collaboration (by a factor smaller than $1.5$ in the whole mass range), 
when applied to the region of DM masses from 10 to 100 GeV with pure annihilation channels. To incorporate the latest results from dSphs~\cite{Ackermann:2015zua} (Pass 8 analysis), we have estimated the improvement of this data set with respect to the previous analysis~\cite{Ackermann:2013yva} in a factor 4 on $\xi^2\sigmav$ for all DM masses.

\subsection{Results}

\begin{figure}[t!]
	\begin{center}  
	 \epsfig{file=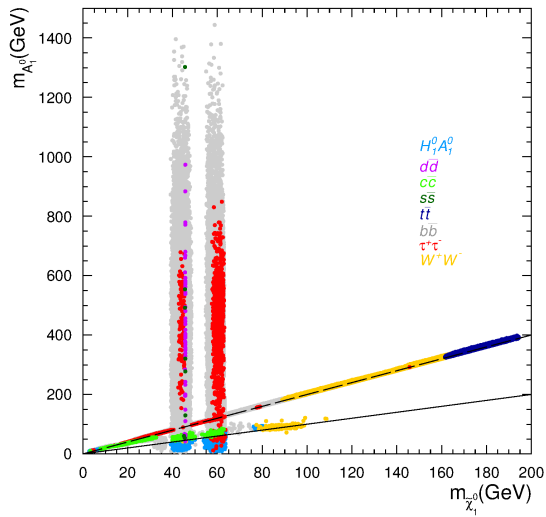,width=7.5cm}
	\end{center}
	\caption{\small Lightest pseudoscalar mass versus the lightest neutralino mass for all the points that satisfy the experimental constraints. The different colours denote the main annihilation channel in the Galactic halo, as indicated in the legend.}
	\label{fig:neut-ma}
\end{figure}

Figure \ref{fig:neut-ma} shows the mass of the lightest pseudoscalar as a function of the mass of the lightest neutralino for all the points in the scan that satisfy the different experimental constraints. 
Not surprisingly, the upper bound on the neutralino relic density is responsible for removing extensive regions of the parameter space.
As we can observe from this figure, the vast majority of the solutions found satisfy the condition for resonant annihilation with either the lightest pseudoscalar (when $\neutlmass \approx 1/2\, \phmassl$, as indicated by the dashed line), or with the $Z$ boson, or with the SM-like CP-even Higgs (the two narrow vertical bands for $\neutlmass\approx 1/2\, \mz$ and $\neutlmass\approx 1/2\, m_{\hsm}$, respectively).
The correct relic density can also be obtained when the annihilation into a pair of very light pseudoscalars is kinematically allowed (below the solid line that corresponds to $\neutlmass = \phmassl$). 
Some points in our scan appear to cluster around the line with $\neutlmass = \phmassl$, but these in fact correspond to cases in which the resonant condition with the second-lightest Higgs is satisfied, when $H_2^0$ is heavier than the SM Higgs.
These conditions are extremely difficult to achieve and have been widely discussed in the literature \cite{Gunion:2005rw,Vasquez:2010ru,Cao:2011re,Vasquez:2012hn,Cao:2013mqa}. Given the low-mass range considered in this analysis, and our choice of the soft masses for sleptons and squarks, no coannihilation effects are present (only a few points present coannihilation with the second-lightest neutralino).

\begin{figure}[t!]
	\begin{center}  
         \epsfig{file=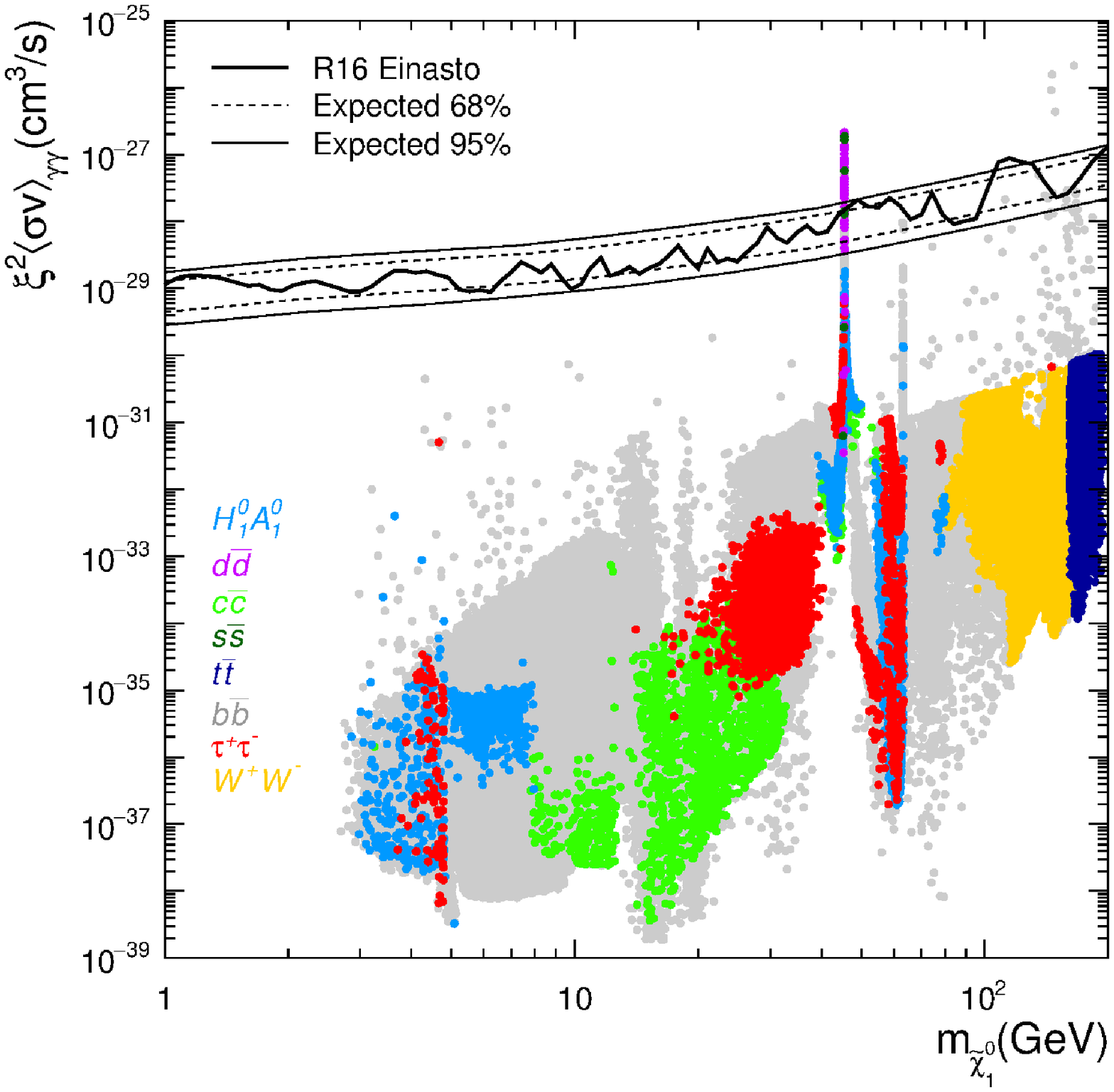,width=7.5cm}
	 \epsfig{file=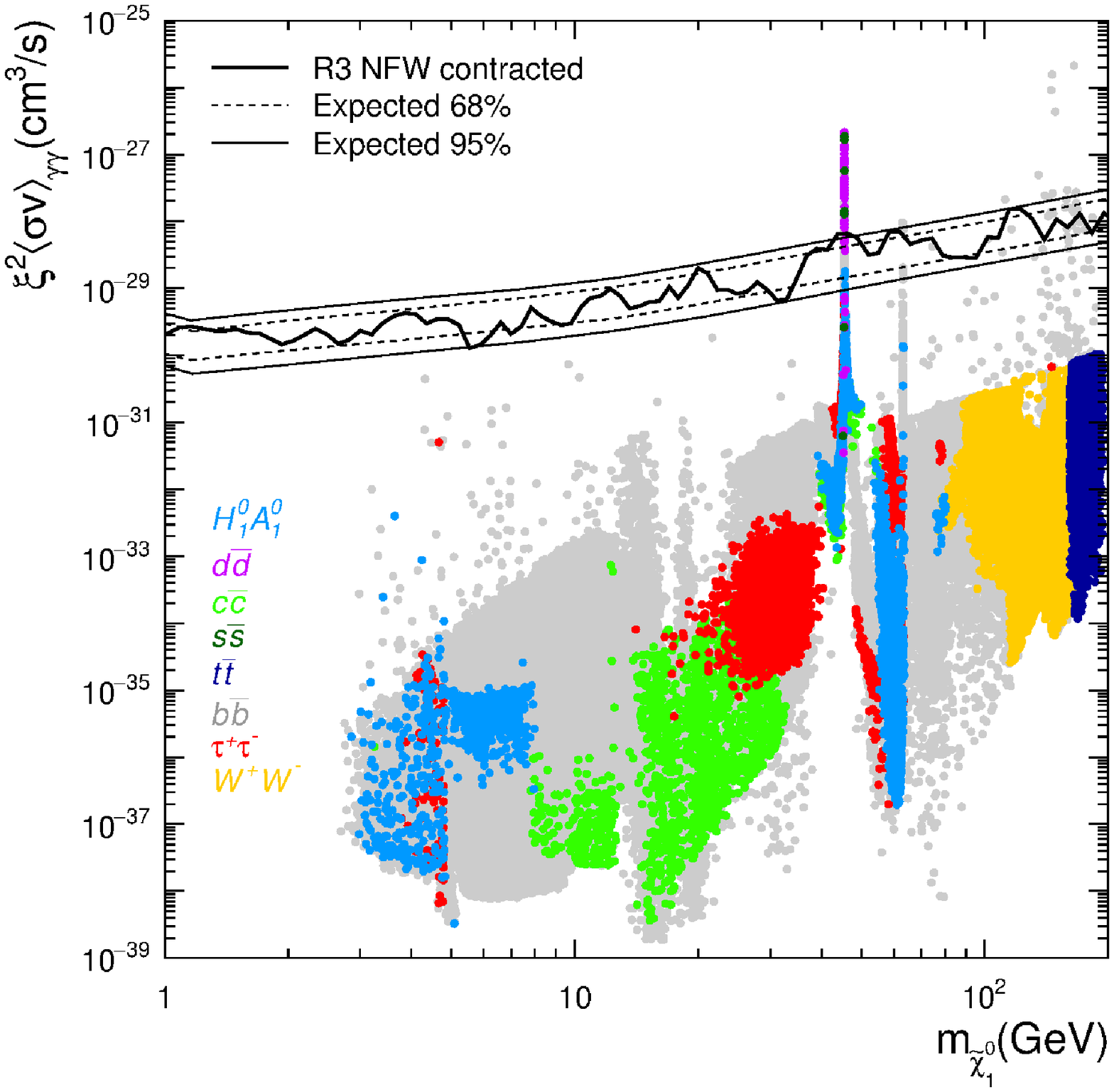,width=7.5cm}
	\end{center}
	\caption{\small Thermally averaged neutralino annihilation cross section into two photons in the Galactic halo as a function of the neutralino mass. 
All the points fulfil the experimental constraints, including bounds from direct detection experiments and Fermi-LAT data on dSphs, and have a relic abundance $0.001<\Omega_{\neutl}h^2<0.13$.
Each colour represents different dominant annihilation channels in the Galactic halo, and the different plots correspond to different assumptions on the halo model.} 
\label{fig:sigvgg_neutralino}
\end{figure}

In Figure \ref{fig:sigvgg_neutralino}, we represent the thermally averaged neutralino annihilation cross section into two photons in the Galactic halo and compare it with the Fermi-LAT upper bounds, derived for the Einasto and NFWc DM density profiles in their corresponding optimised ROIs \cite{Ackermann:2015lka}. Although the results from Ref.\,\cite{Albert:2014hwa} extend the bounds to lower masses for the Einasto profile, they do not exclude any new points in our scan.
The Fermi-LAT bounds are shown as solid black lines with their corresponding 68\% and 95\%CL expectations (black dotted and thin black lines respectively). 
Note that our theoretical predictions are in general very far from the current constraints. Nevertheless, large values of ${\sigmavgg}$ can be obtained in the vicinity of the resonances due to the well-known Breit-Wigner
enhancement, which has been studied in various models \cite{Feldman:2008xs,Ibe:2008ye,Guo:2009aj,AlbornozVasquez:2011js,Chatterjee:2014bva}.
The most obvious resonance is the one with the $Z$ boson, which appears as a prominent peak for $\neutlmass\approx 45$~GeV. The resonance with the lightest pseudoscalar or with a CP-even Higgs is not restricted to a single value of neutralino mass: due to the variation in the Higgs mass we can observe points with an increased $\sigmavgg$ for the whole mass range (the peak for $\neutlmass\approx 62$~GeV corresponds to the resonance with the SM-like Higgs). 
Notice that the conditions for a Breit-Wigner enhancement are more easily satisfied in the $Z$ boson resonance, given its larger decay width, whereas for Higgs boson resonances this only happens for an extremely small range of masses.
In the neutralino case
we must therefore conclude that the observation of gamma-ray lines by the Fermi-LAT satellite is very unlikely, and subject to exceptional conditions.

A small fraction of our solutions, with $\tau\bar \tau$ final states, feature stau exchange, and therefore potential contribution from VIB. This occurs for some of the red points in Figure 2 with $m_\chi\approx 45$~GeV and $m_\chi\approx 60$ GeV. 
In Ref.\,\cite{Bringmann:2012vr}, it is shown that when $\mu_{\tau}\equiv \left(m_{\tilde\tau}/{m_\chi}\right)^2$ is large, the VIB contribution scales as $\mu_{\tau}^{-4}$.
In our scan, the mass of the neutralino is much smaller than the mass of the stau and  $\mu_\tau>4$, thus VIB is not expected to be relevant.

\begin{figure}[t!]
	\begin{center}  
     \epsfig{file=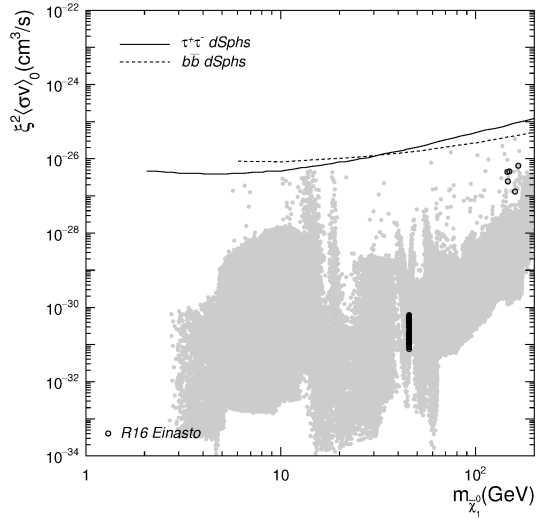,width=7.5cm}
     	\end{center}
\caption{\small Velocity-averaged annihilation cross section of neutralinos in the Galactic halo as a function of the neutralino mass. 
All the experimental constraints, including bounds from direct detection experiments and Fermi-LAT dSph data have been taken into account. 
The solid and dotted lines correspond to the Fermi-LAT upper bounds from dSphs, assuming annihilation into $\tau^+\tau^-$ or $b\bar b$, respectively.
Black dots correspond to those that exceed Fermi-LAT constraints on gamma-ray lines in Figure \ref{fig:sigvgg_neutralino} for the Einasto profile.}
\label{fig:id_neutralino}
\end{figure}

It is illustrative to compare these results with other strategies for indirect DM searches.
In Figure \ref{fig:id_neutralino}, we show the total neutralino annihilation cross section in the Galactic halo as a function of the neutralino mass 
for the points that fulfil all the experimental limits, including the Fermi-LAT constraint from continuum gamma-ray emission from dSphs, currently the most stringent bound on $\sigmav$ for the range of DM masses considered in this work. 
As a reference, we have depicted the Fermi-LAT upper bounds from dSphs, assuming annihilation into $\tau^+\tau^-$ or $b\bar b$ (solid and dotted lines respectively). 
Black dots in this plot correspond to the few points that exceed the Fermi-LAT bound in Figure \ref{fig:sigvgg_neutralino}. For concreteness we have chosen the Einasto profile but the results are virtually insensitive to the halo choice.
As pointed out in Refs.\,\cite{Ferrer:2006hy,AlbornozVasquez:2011js,Lee:2012bq,Das:2012ys,Chalons:2012xf}, the points with a larger $\sigmav$ generally correspond to those with resonant annihilation through the lightest pseudoscalar. Note that these regions are significantly less populated in our scan, the main reason being the new bounds on the Higgs sector (very light pseudoscalars can contribute significantly to the invisible SM Higgs branching ratio \cite{Cao:2011re,Vasquez:2012hn}) and the improvement in the determination of low-energy observables (with the measurement of BR($\bmumu$) playing an important role).
We should emphasize at this point that the resonant condition for annihilation in the DM halo is satisfied for a much smaller mass range than in the early Universe, given the much smaller WIMP velocity dispersion in the halo. For this reason, the predicted $\xi^2\sigmav$ is in general very small.

Notice that, in principle, one can also have box-shaped features \cite{Ibarra:2012dw,Ibarra:2013eda} in the gamma-ray spectrum for final states that contain scalar or pseudoscalar Higgses (which can decay in-flight to a pair of photons). However, comparing Figures \ref{fig:sigvgg_neutralino} and \ref{fig:id_neutralino} we observe that the total annihilation cross section for $\higgsl\phiggsl$ is very small (of the order of $\sigmav\lesssim10^{-28}$~cm$^3$\,s$^{-1}$ for points outside the $Z$ resonance), suggesting that the resulting contribution to a box-shaped feature is several orders of magnitude below the current Fermi-LAT sensitivity for line searches. 
It is also worth noting that variations in the soft slepton mass, $m_{L.E}$, and lepton trilinear coupling, $A_E$, might lead to points in which the stau mass is similar to that of the neutralino. Given the current lower bounds on the stau mass, this is only possible for neutralino masses above $\sim 100$~GeV. Even though we do not expect an increase in the contribution to gamma-ray lines, these points might lead to observable features due to VIB \cite{Bringmann:2007nk}. Similarly, a VIB enhancement can occur in the $W^+W^-$ final state \cite{Bringmann:2007nk} when neutralinos and charginos are almost degenerate, although it has been argued that this enhancement is more prominent in the limit of large neutralino masses \cite{Cannoni:2010my}.

\begin{figure}[t!]
	\begin{center} 
     \epsfig{file=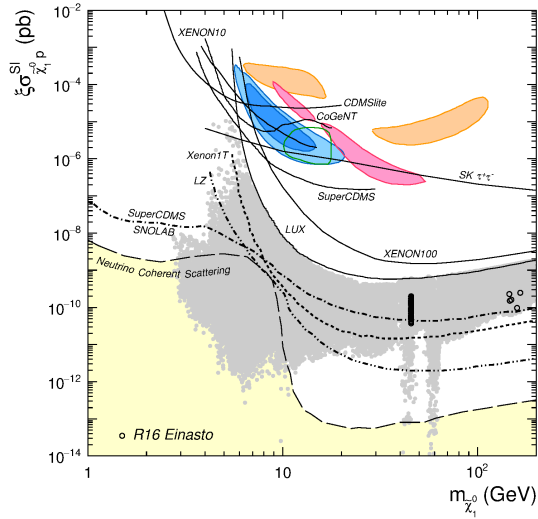,width=7.5cm}
	\end{center}
	\caption{\small Spin-independent neutralino-proton 
	cross section as a function of the neutralino mass. 
	All the experimental constraints, including bounds from direct detection experiments and Fermi-LAT data on dSphs have been considered. Solid lines represent the current  experimental upper bounds from direct detection experiments, whereas dotted lines are the projected sensitivities of next-generation detectors. Both of them correspond to a rescaled SHM. Black circles correspond to points whose $\xi^2\langle\sigma v\rangle_{\gamma\gamma}$ exceeds the Fermi-LAT bounds for the Einasto profile.
	}
\label{fig:dd_neutralino}
\end{figure}

Let us now put these results in a wider context and contrast them with direct detection techniques.
In Figure \ref{fig:dd_neutralino}, we display the spin-independent neutralino nucleon scattering cross section, $\sineut$, as a function of its mass. 
Direct detection constraints are implemented using the procedure of Ref.\,\cite{Marcos:2015dza} from SuperCDMS and LUX results.
The upper bounds of all the direct detection experiments for a Standard Halo Model (SHM) profile and only SI interactions are shown for comparison by means of solid lines. As a reference, we also show the limits for CRESST \cite{Angloher:2014myn}, and SuperKamiokande \cite{Choi:2015ara} (the latter is extracted assuming annihilation into $\tau^+\tau^-$). 
Interestingly, the region of low mass neutralinos (with masses as small as 3~GeV) is still viable and the predicted $\sineut$ for a large fraction of the points found lies within the reach of second generation experiments, such as SuperCDMS \cite{Brink:2012zza} and LZ \cite{Malling:2011va}, whose projected sensitivities are shown by means of dashed lines\footnote{We remind the reader that all the constraints and prospects have been rescaled according to a local DM density of $0.4$~GeV$/$cm$^3$.}. 
Notice, however, that a substantial region of the explored parameter space lies below the predicted bound where coherent neutrino scattering becomes a background for these searches. 
Our results for $\sineut$ versus $\neutlmass$ are comparable to those obtained in Ref.\,\cite{Huang:2014cla} (which were obtained to fit the GCE), but span a larger range of solutions.

\begin{figure}[t!]
	\begin{center} 
     \epsfig{file=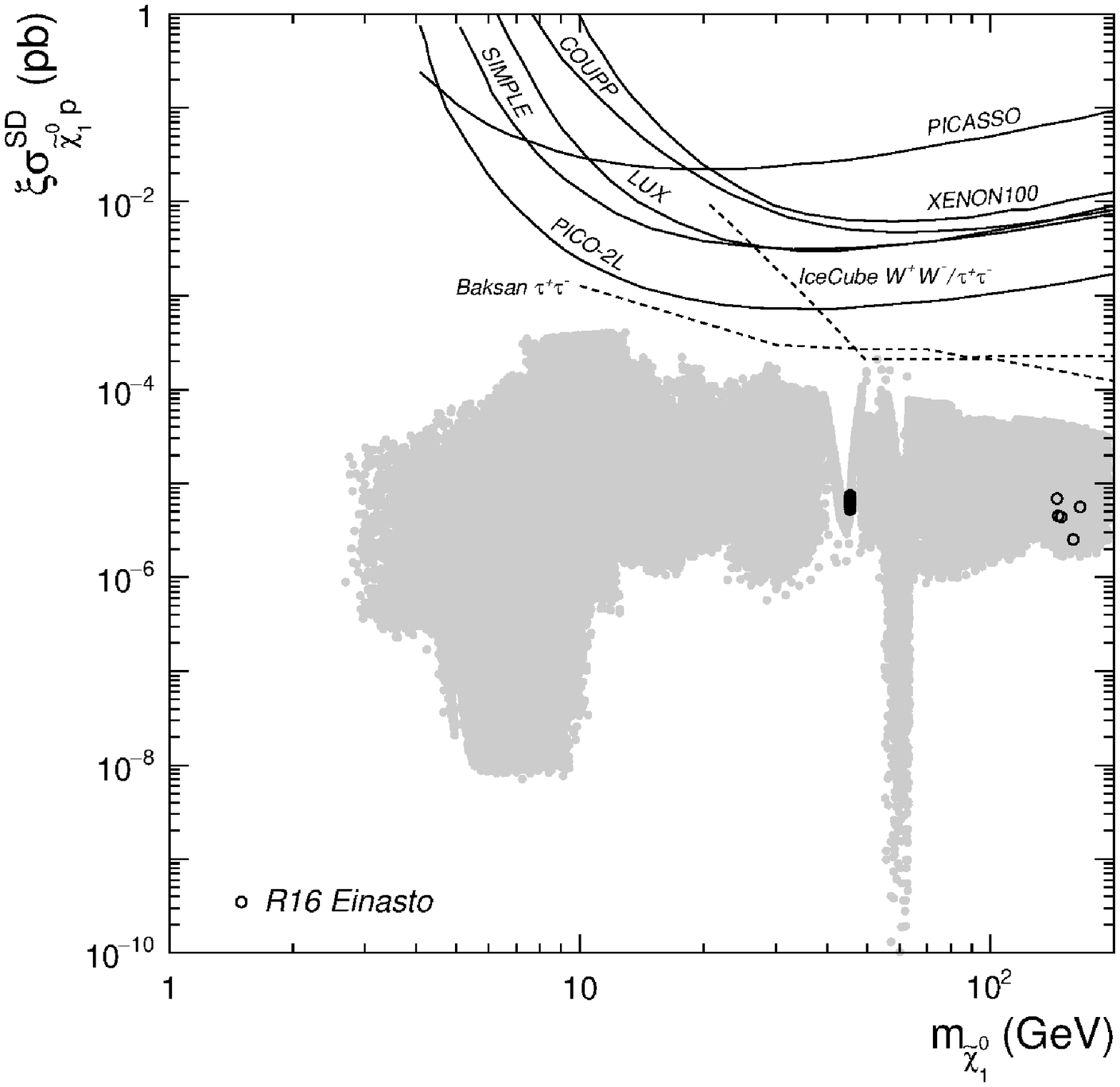,width=7.5cm}
     \epsfig{file=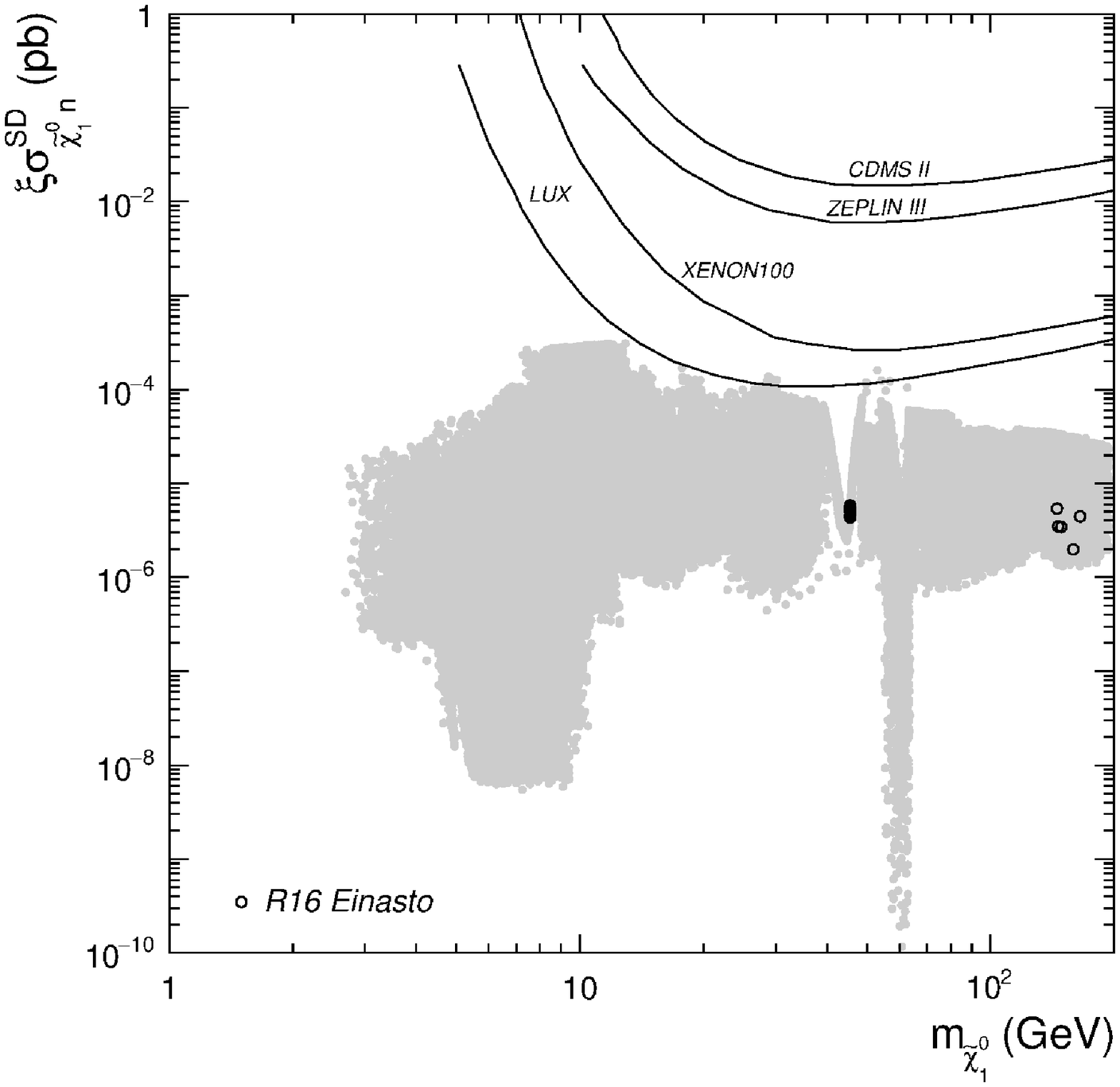,width=7.5cm}\\ 
     \epsfig{file=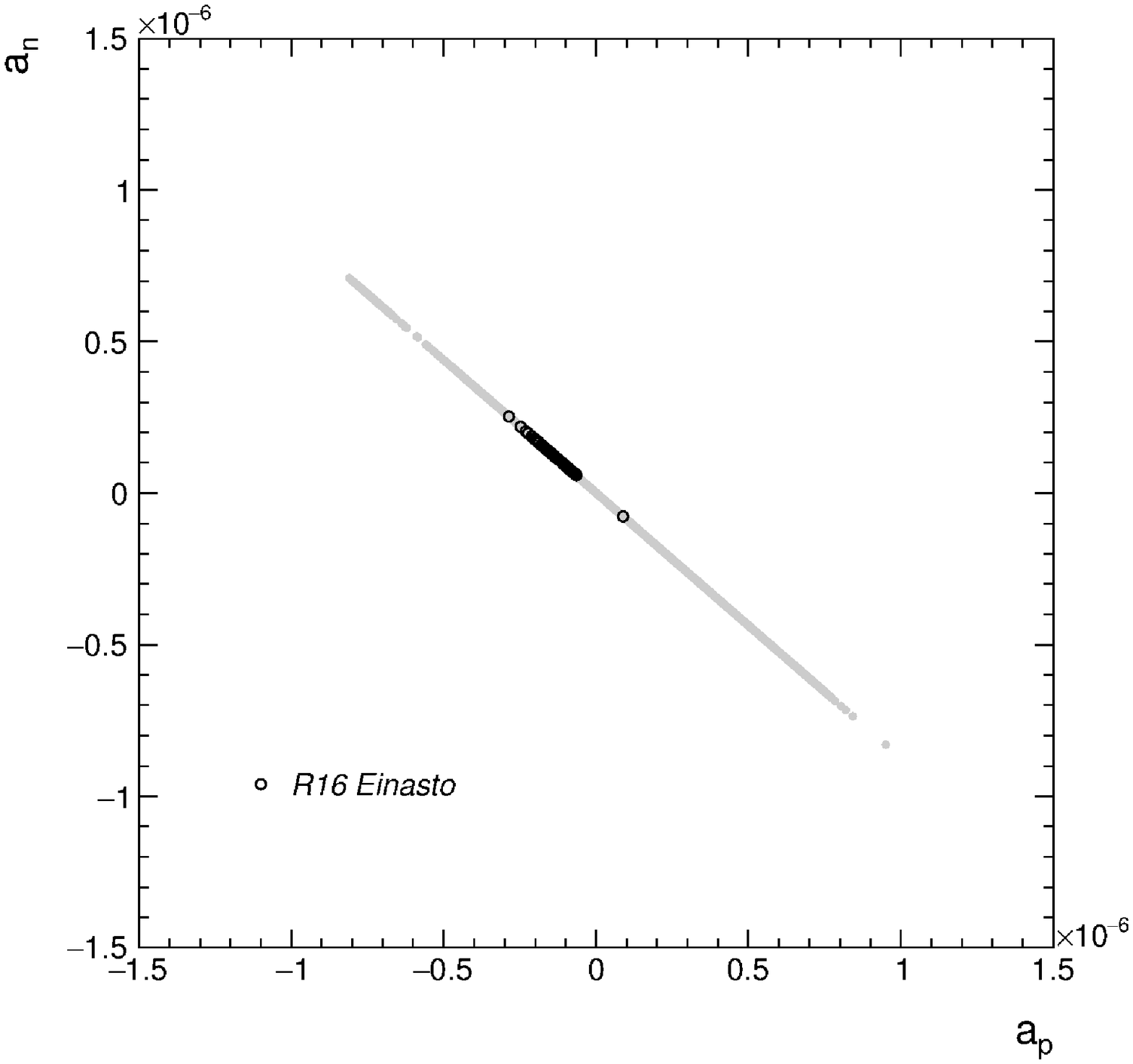,width=7.5cm}
     \epsfig{file=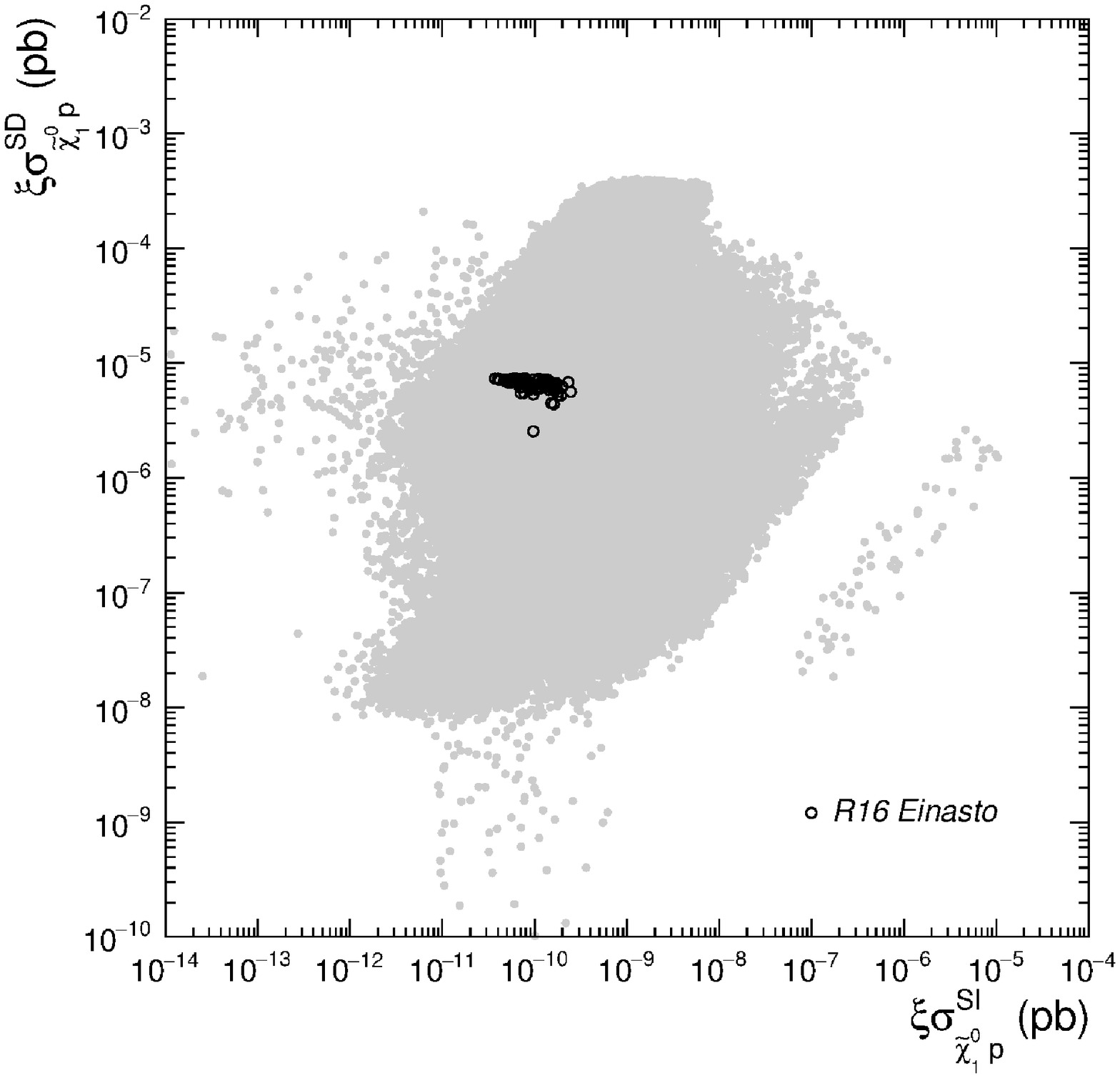,width=7.5cm}
	\end{center}
	\caption{\small Top row: spin-dependent neutralino-proton (left) and spin-dependent neutralino-neutron (right) cross sections as a function of the neutralino mass.
	Lower row: $a_n$ and $a_p$ coefficients for the spin-dependent cross section (left) and spin-dependent versus spin-independent neutralino-proton scattering cross section. 
	Solid lines represent the limits from direct detection experiments for a rescaled SHM. 
	 All the experimental constraints, including bounds from direct detection experiments and Fermi-LAT data on dSphs have been considered. 
	 No colour code is use to distinguish different annihilation channels and black dots represent points that exceed Fermi-LAT constraints on gamma-ray lines in Figure \ref{fig:sigvgg_neutralino} for the Einasto profile.}
\label{fig:sddd_neutralino}
\end{figure}

Regarding the results for the spin-dependent contribution with protons, $\sdpneut$, and neutrons $\sdnneut$, in general they are rather small, below the current upper bounds derived by PICO-2L \cite{Amole:2015lsj} (which dominates for SD-proton) and LUX  (which supersedes Xenon100 \cite{Aprile:2013doa} for SD neutron). We show these results in the upper row of Figure \ref{fig:sddd_neutralino}.
In these plots, we also quote published bounds on $\sigma^{SI}_p$ from the neutrino telescopes, SuperKamiokande \cite{Choi:2015ara} IceCube \cite{Aartsen:2012kia}, BAKSAN \cite{Boliev:2013ai}, and ANTARES \cite{Zornoza:2014cra}, which produce competitive results for DM particles which annihilate preferentially into $\tau^+\tau^-$ or $W^+W^-$.
Remarkably, a small (but non-negligible) part of the scanned parameter space has been excluded based on the sensitivity of LUX spin-dependent cross section with neutrons. 
The plot on the bottom left of Figure \ref{fig:sddd_neutralino} shows the spin-dependent coupling of the lightest neutralino to neutrons, $a_n$, and protons, $a_p$. As we can observe, there is a perfect correlation between both quantities ($a_n\approx - 0.8 \, a_p$). This is a result of the different quark composition of neutrons and protons and the fact that in this case the dominant contribution to spin-dependent interactions is the exchange of a $Z$ boson (squarks are heavy in our scan).

Finally, the plot on the bottom right of  Figure \ref{fig:id_neutralino} represents the spin-dependent versus spin-independent cross sections (with protons in both cases). This figure can be understood as a reanalysis of low-mass neutralino dark matter in the NMSSM with updated constraints. 
Since the direct detection constraints are a function of the DM mass, it is difficult to visualize an exclusion line in this plot. These results perfectly exemplify the necessity of exploring this parameter space with complementary targets which are sensitive to both contributions.

To summarise, the constraints on monochromatic gamma-ray emission for the case of neutralino DM in the NMSSM are in general less important than the dSph bounds and direct detection limits. Only in the narrow resonances (mainly with the $Z$ boson) we have found a reduced number of points for which $\sigmavgg$ exceeds the current constraints.


\section{Spectral features from RH sneutrino annihilation}
\label{sec:sneutrino}

The NMSSM can be extended with RH neutrinos in an attempt to accommodate a see-saw mechanism that explains the smallness of neutrino masses \cite{Kitano:1999qb,Deppisch:2008bp}. To this aim, in Ref.\cite{Cerdeno:2008ep} a gauge singlet RH neutrino superfield, $N$, that couples to the superfield $S$ was included in the NMSSM superpotential. In this construction a Majorana mass term for the RH neutrinos is generated after radiative Electroweak symmetry breaking, which is naturally of the order of the EW scale. A see-saw mechanism with an EW scale Majorana mass generally results in small values for the neutrino Yukawa coupling. This, in turn, implies that in this construction the left and right-handed fields have a negligible mixing. It was previously shown that the lightest RH sneutrino, $\snr_1$, can be a viable DM candidate in wide areas of the parameter space \cite{Cerdeno:2008ep,Cerdeno:2009dv}, including the low-mass range \cite{Cerdeno:2011qv,Cerdeno:2014cda}, and might also account for the GCE \cite{Cerdeno:2015ega}.

This model includes five new parameters. Three of them, namely the soft mass parameter of the RH sneutrino field, $\mn$, the coupling between the RH sneutrino and the singlet Higgs, $\lambda_N$, and the corresponding trilinear term in the Lagrangian, $A_{\lambda_N}$, are related to the mass and couplings of the RH sneutrino and determine its phenomenological properties as a DM candidate. The other two parameters: the neutrino Yukawa coupling and trilinear parameter are in general small and have a negligible impact on the DM sector. The RH sneutrino can annihilate into a wide range of final states. As in the case of the neutralino, the presence of light scalar and pseudoscalar Higgs bosons is very beneficial to reproduce the correct relic abundance while fulfilling the LHC constraints on the SM-like Higgs couplings.

The RH sneutrino in the NMSSM can give rise to a complex spectrum, displaying lines and box-shaped spectral features\footnote{For a more detailed discussion on these features see Refs.~\cite{Ibarra:2012dw,Ibarra:2013eda}.}.
Regarding the monochromatic gamma-ray emission from RH sneutrino annihilation, the only final states that contain photons at the one-loop level are $\gamma\gamma$ and $Z\gamma$ 
(as the $H\gamma$ final state violates spin conservation). 
The RH sneutrino interacts with SM particles through the Higgs sector, with a coupling 
strength controlled by the parameter $\lambda_N$. 
This leaves two possibilities for diphoton final states in RH sneutrino annihilation: $s$-channel CP-even Higgs, $\higgsi$, exchange (with a loop of electrically charged fermions, sfermions or bosons), and direct annihilation through a loop of charged Higgses, $\chiggspm$.
The Feynman diagrams for these processes are depicted in Figure \ref{fig:diphotonann}, and for reference, the corresponding matrix elements are given in Appendix~\ref{sec:oneloopcalc}.

\begin{figure}[!t]
\begin{center}
\includegraphics[scale=0.5]{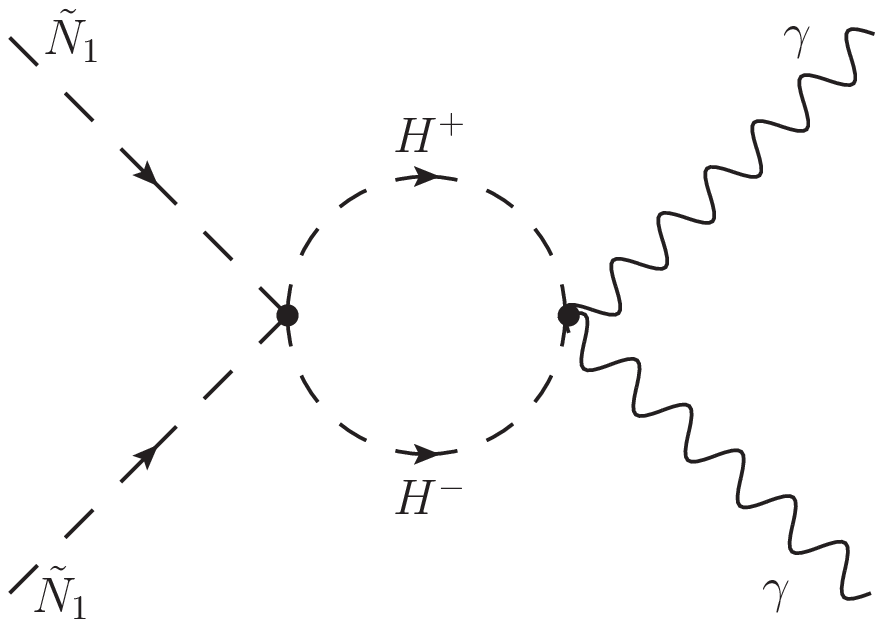}\quad\quad\quad\quad
\includegraphics[scale=0.5]{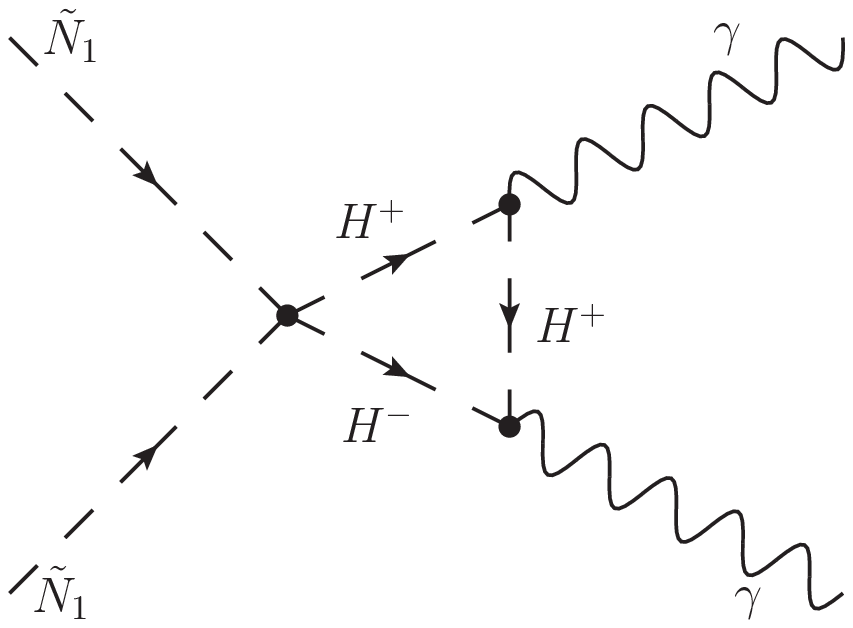}\\[1ex]
\includegraphics[scale=0.5]{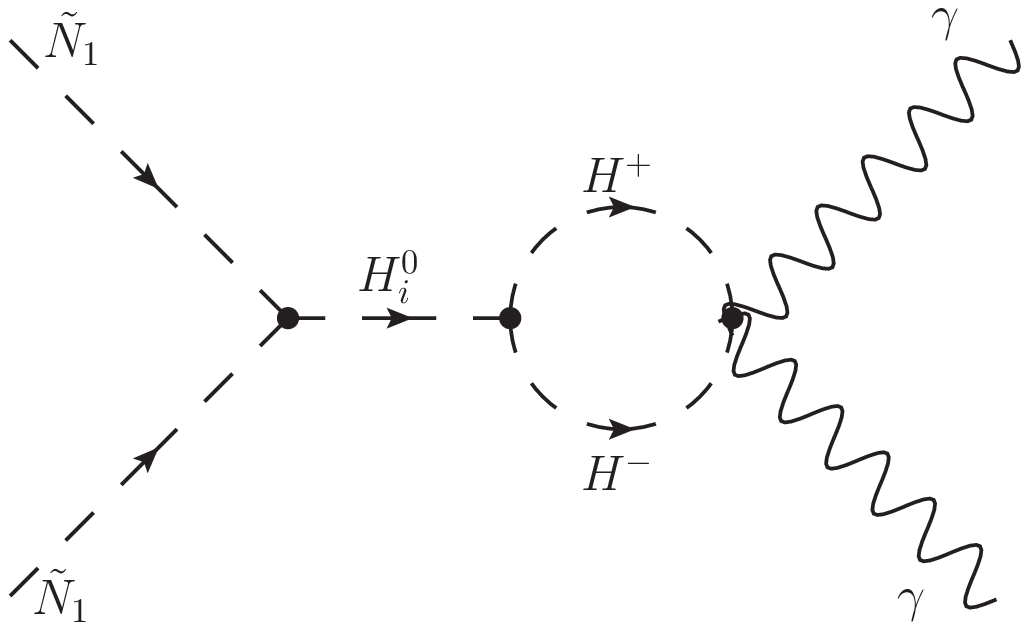}\quad\quad
\includegraphics[scale=0.5]{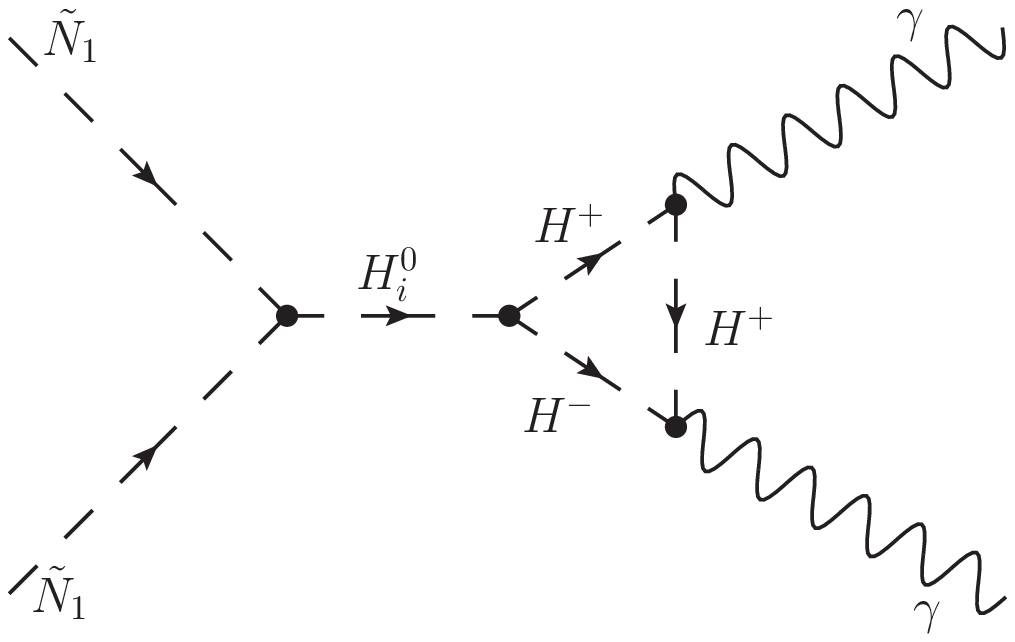}\\[1ex]
\includegraphics[scale=0.5]{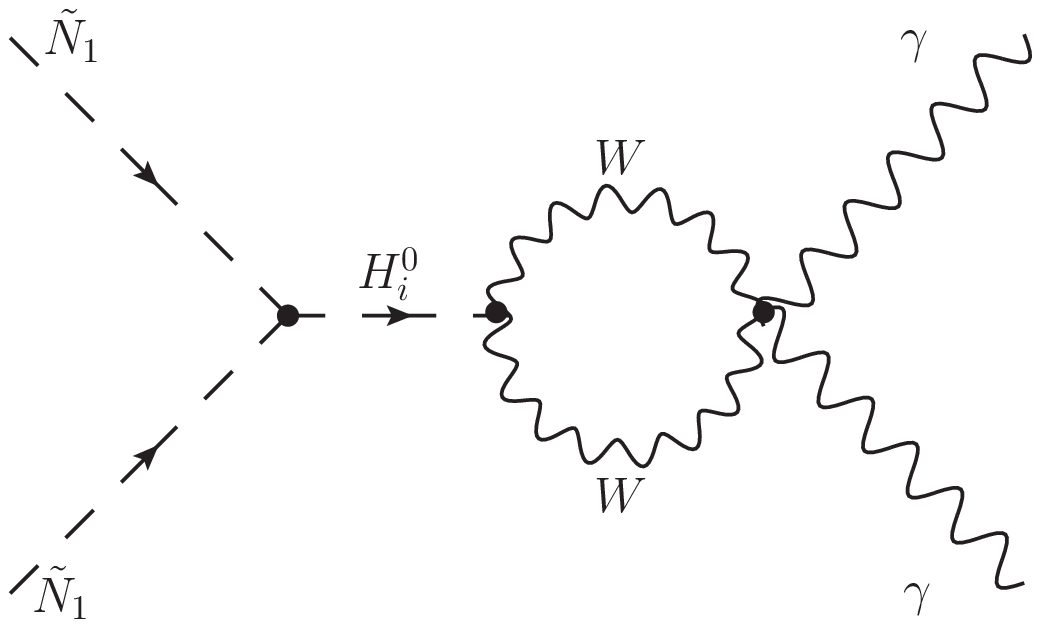}\quad\quad
\includegraphics[scale=0.5]{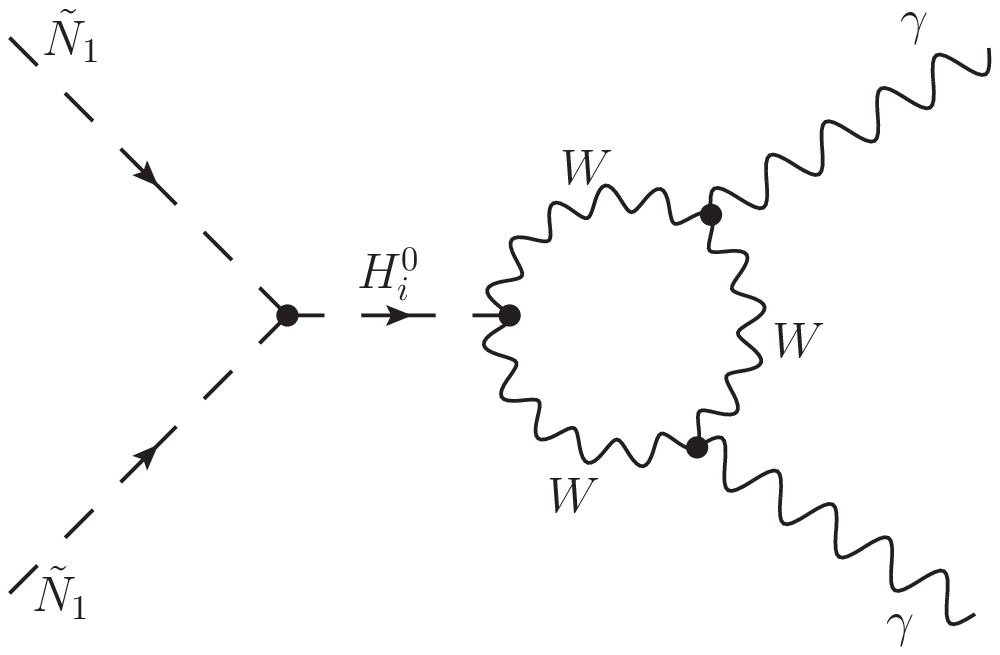}\\[1ex]
\includegraphics[scale=0.5]{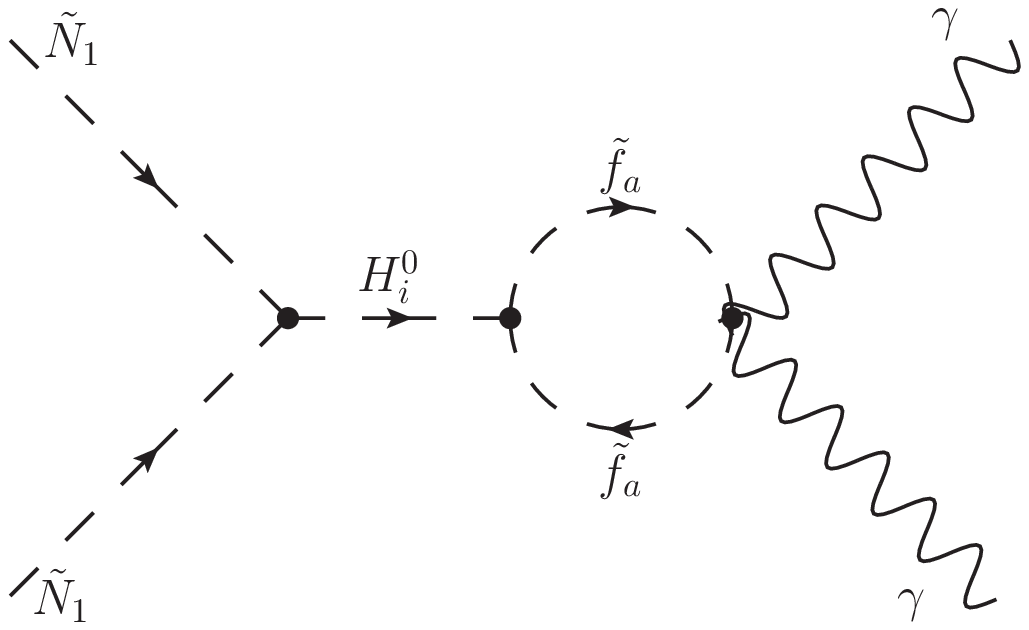}\quad\quad
\includegraphics[scale=0.5]{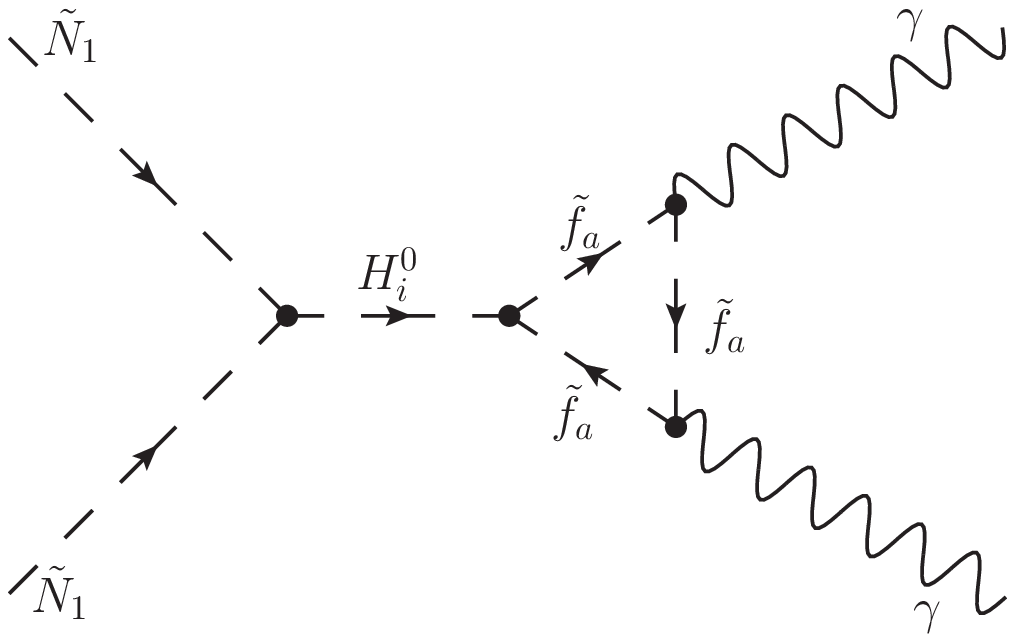}\\[1ex]
\includegraphics[scale=0.5]{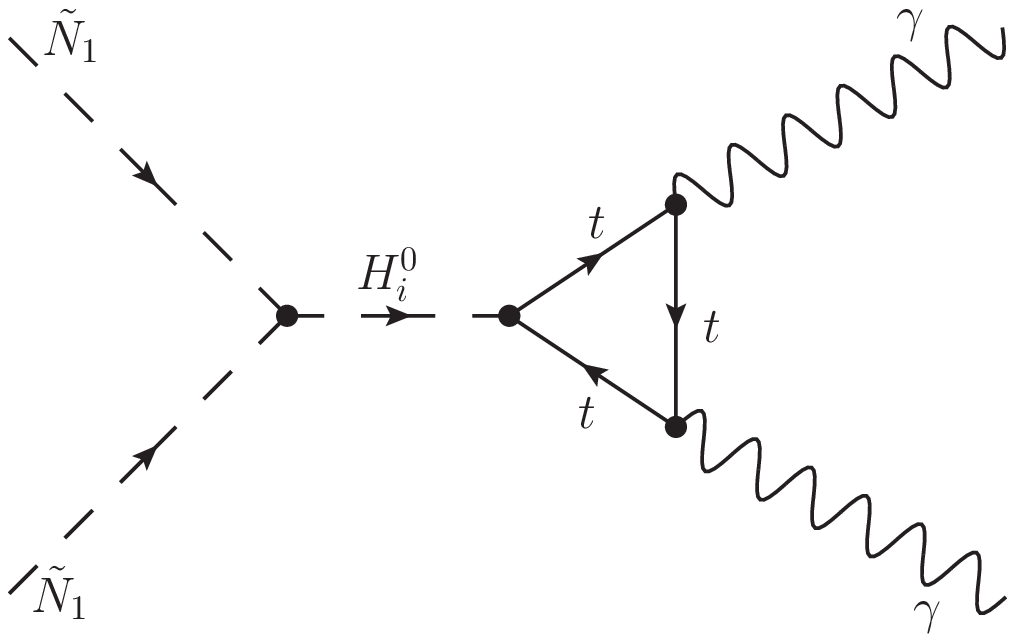}\quad\quad
\includegraphics[scale=0.5]{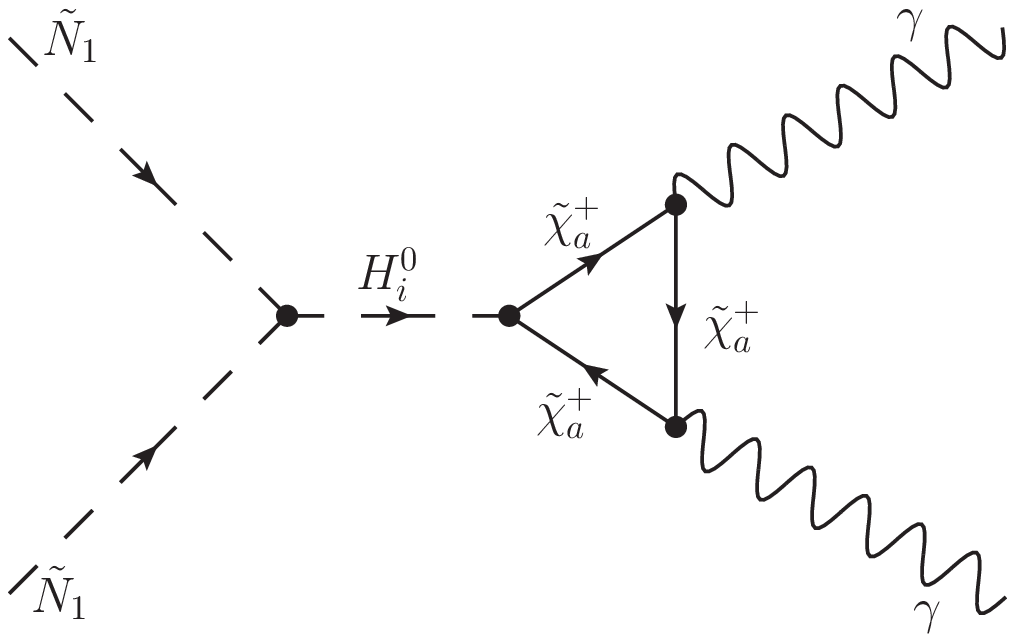}
\end{center}
\caption{\small Contributions to the RH sneutrino annihilation into a pair of photons, through a $\chiggspm$ loop or mediated by a $\higgsi$ and a loop of SM and NMSSM particles.}
\label{fig:diphotonann}
\end{figure}

Concerning box-shaped contributions to the spectrum, in Ref.\,\cite{Cerdeno:2015ega}, we showed how these features were obtained from RH sneutrino annihilation into pairs of boosted light scalar and pseudoscalar Higgs bosons, which subsequently decay in flight.
In order to incorporate bounds on these features, we have derived the corresponding limits on the annihilation cross section from the Fermi-LAT gamma-ray line 
bounds.
Afterwards we have applied these limits to our prediction weighted by the fractional DM density squared, $\xi^2$, along the box width.

\subsection{Details of the scan and experimental constraints}

We have implemented the RH sneutrino construction in {\tt CalcHEP 3.4.3}~\cite{Belyaev:2012qa} model files, including the one-loop matrix elements for photon final states in RH sneutrino annihilation. We define the input parameters at the EW scale and the scan procedure (using {\tt micrOMEGAs 3.6.9} and  {\tt MultiNest 3.0}) and 
experimental constraints are implemented in the same way as we have described for the case of the neutralino in Section\,\ref{sec:neutscan}\footnote{As in the previous section, it should be noted that we do not attempt to fit the GCE.}.  
Due to the smallness of the neutrino Yukawa coupling there is no extra contribution from the RH sneutrino sector to any of the low-energy observables.

We have scanned the same ranges of the parameter space as in Refs.\,\cite{Cerdeno:2014cda,Cerdeno:2015ega}, with values of the RH sneutrino mass up to 150~GeV. 
For our choice of soft terms, the lightest neutralino becomes the lightest supersymmetric particle for larger values of the RH sneutrino mass. 
Table \ref{tab:scan_rhnmssm} summarises the values taken for the different input parameters.
Gaugino soft masses are taken to be $M_1=350~\GeV$, $M_2=700~\GeV$ and $M_3=2100~\GeV$, thus satisfying the Grand Unification relation. Slepton and squark soft masses are equal for the three families, 
$m_{\widetilde{L}}=m_{\widetilde{e}^c}=300~\GeV$, and
$m_{\widetilde{Q}}=m_{\widetilde{u}^c}=m_{\widetilde{d}^c}=1500~\GeV$.
Trilinear soft terms are chosen to be $A_{t}=3700~\GeV$, $A_{b}=2000~\GeV$, $A_{\tau}=-1000~\GeV$.
All these parameters are defined at the EW scale.

\begin{table}[!t]
  \begin{center}
    \begin{tabular}{|c|ccc|}
      \hline
      Parameter & & Range & \\
      \hline
      \hline
      $\tanb$ & & $[4 , 10]$, $[10 , 20]$  & \\
      $\lambda$& & $[0.1, 0.6]$& \\
      $\kappa$& & $[0.01, 0.1]$ & \\
      $\al$& & $[500 , 1100]$ & \\
      $\ak$& & $[-50 , 50]$ & \\
      $\mu$& & $[110 , 250]$ & \\
      $\ln$& & $[0.07 , 0.4]$ & \\
      $\aln$& & $[-1100 , -500]$ & \\
      $\snmassr$& &$[1 , 150]$ & \\     
      \hline
    \end{tabular}
    \vspace{0.4cm}
    \caption{\small Input parameters for the series
      of scans used in this work in the NMSSM with RH sneutrinos. Masses and trilinear parameters are given in GeV. All parameters are defined at the EW scale. 
}
    \label{tab:scan_rhnmssm}
  \end{center}
\end{table}

\subsection{Results}

In Ref.\,\cite{Cerdeno:2014cda}, we studied the phenomenology of this construction and performed a similar scan (limited to lower RH sneutrino masses). From that analysis, we concluded that since the new parameters in the RH sneutrino sector can be varied without affecting the Higgs masses and couplings, it is much easier to reproduce the correct DM relic density than in the neutralino case presented in the previous section. Viable points of the parameter space include resonances with the lightest scalar Higgs (we assume no CP violation in the sneutrino sector and thus no direct coupling to the pseudoscalar Higgs), but are not limited to them. In particular, annihilation into a pair of very light scalar or pseudoscalar Higgses is dominant when these channels are kinematically open. We refer the reader to Figures 1 and 2 of Ref.\,\cite{Cerdeno:2014cda}, where the masses of the lightest scalar and pseudoscalar Higgs bosons are represented as a function of the RH sneutrino mass.

\begin{figure}[t!]
	\begin{center}  
	\epsfig{file=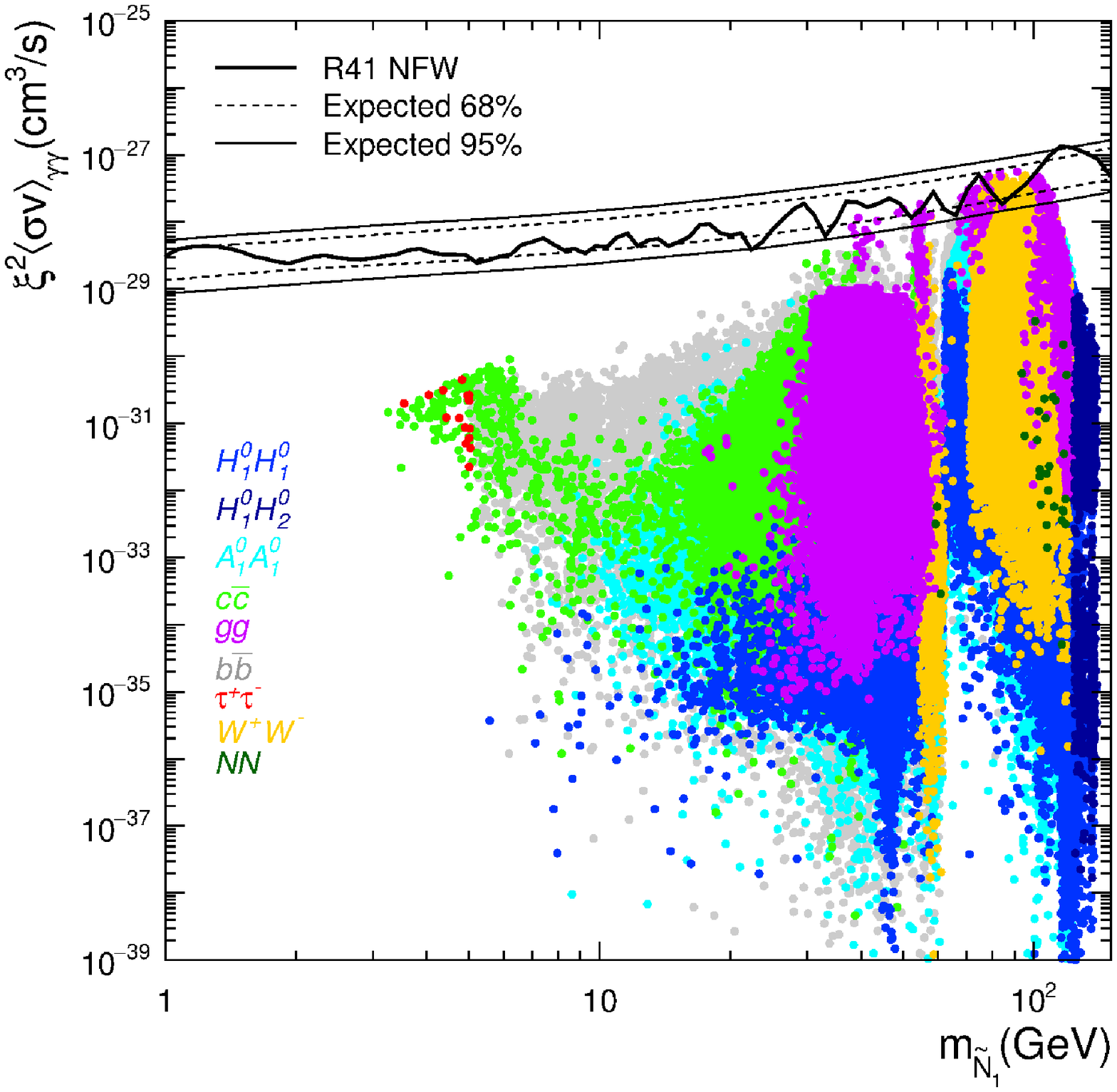,width=7.5cm}
	\epsfig{file=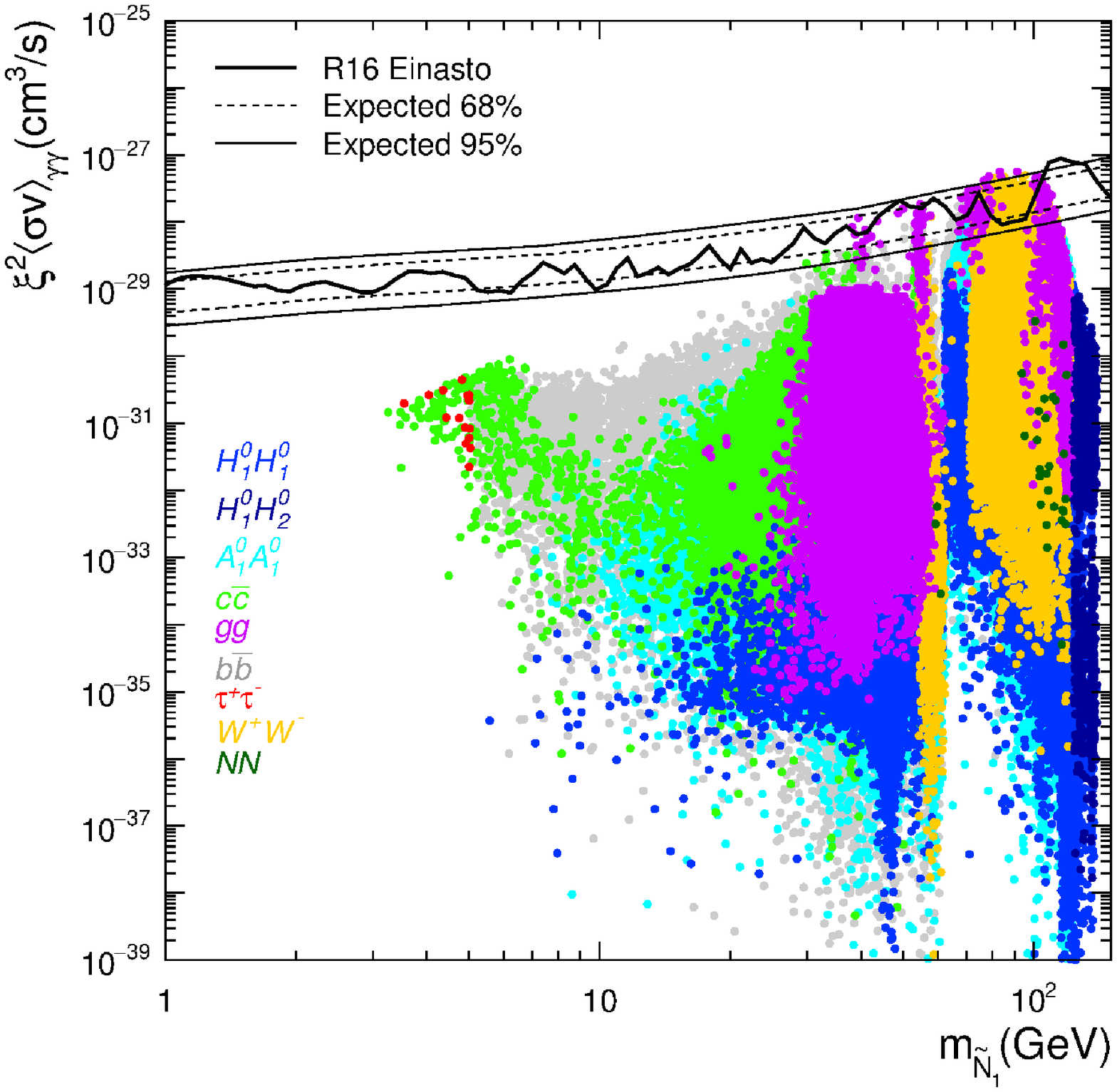,width=7.5cm}\\
	\epsfig{file=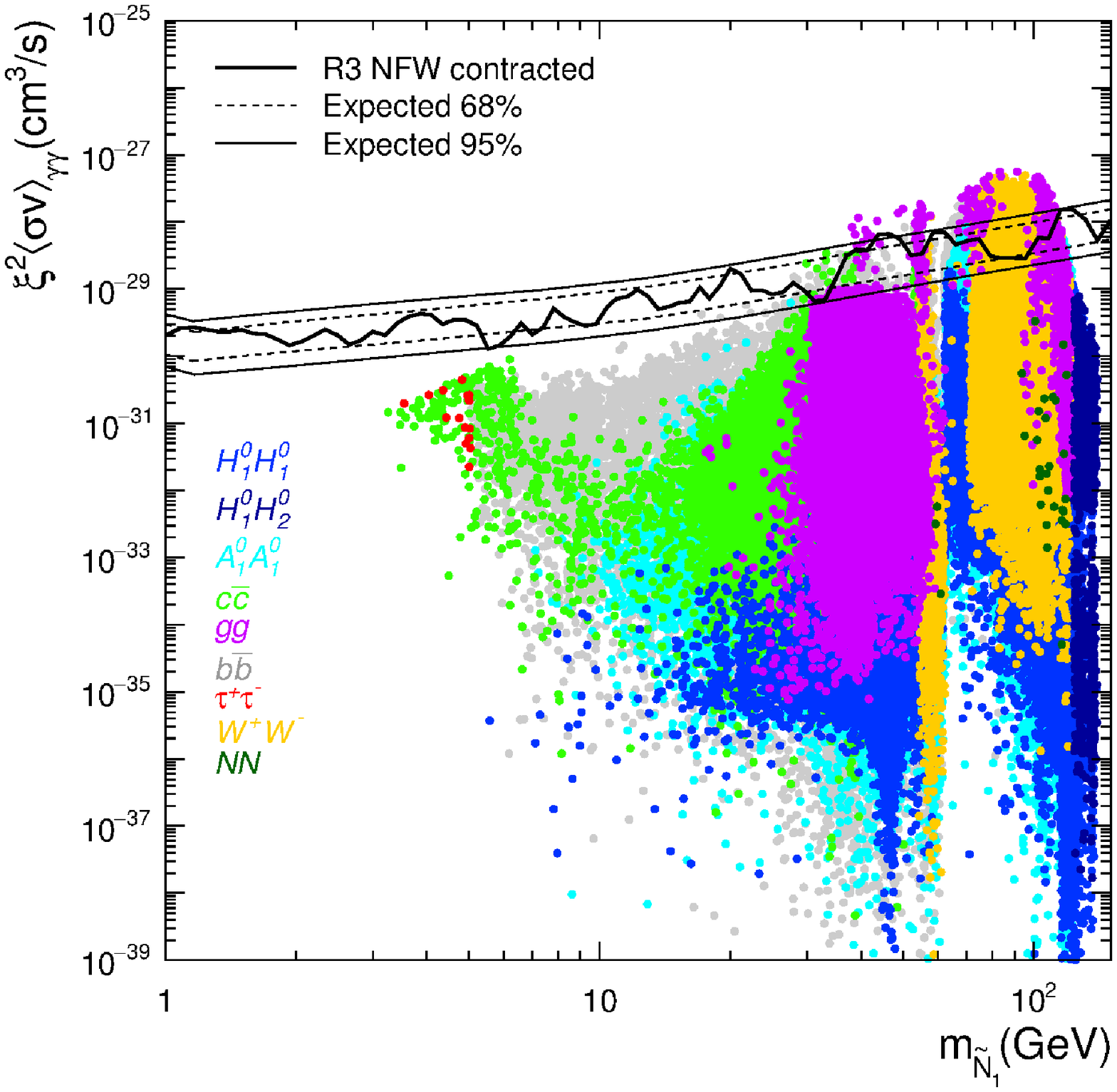,width=7.5cm}
	\epsfig{file=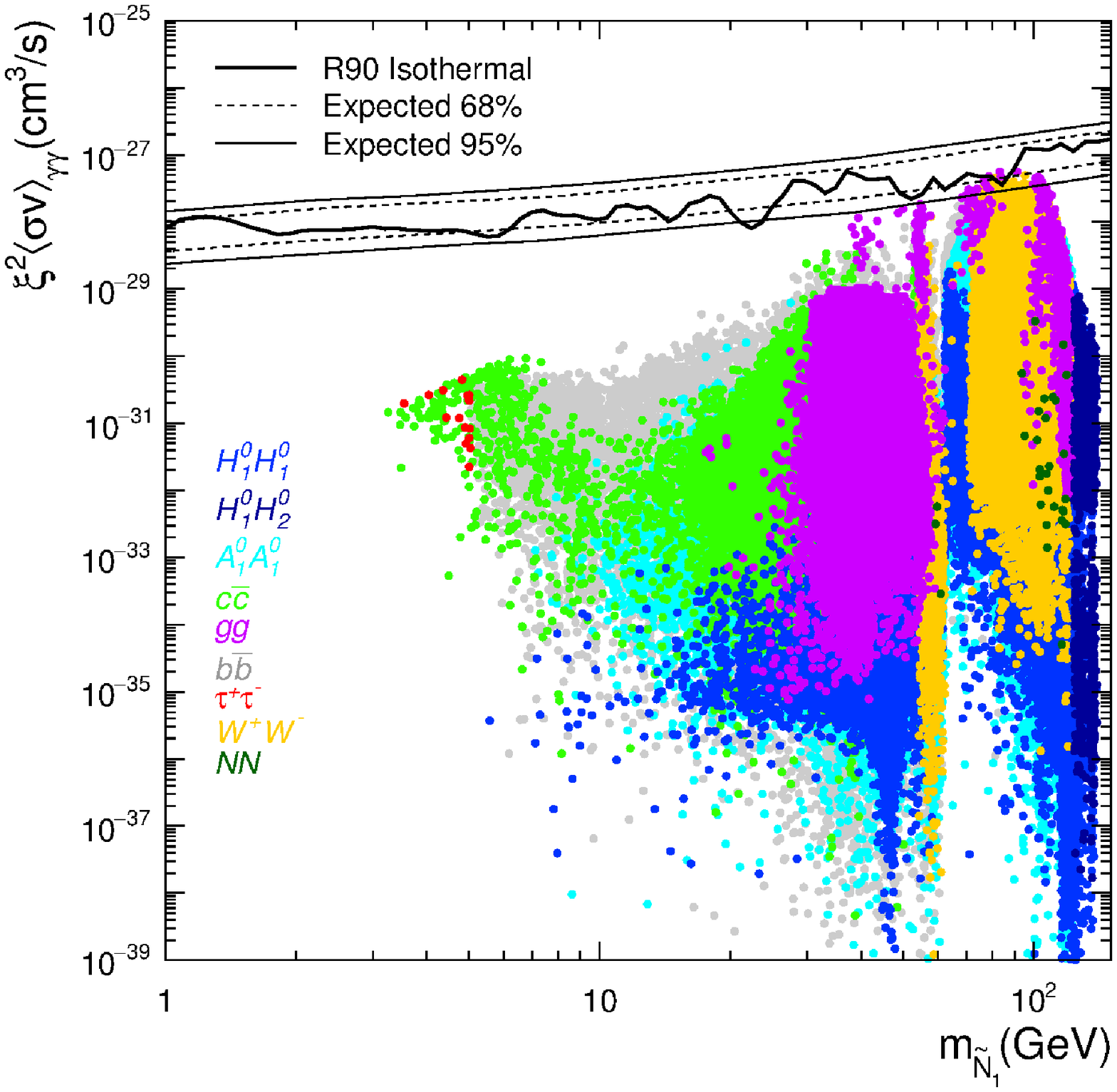,width=7.5cm}
	\end{center}
\caption{\small Thermally averaged RH sneutrino annihilation cross section into two photons in the Galactic halo versus the sneutrino mass. 
All the points fulfil the experimental constraints, including bounds from direct detection experiments and Fermi-LAT data on dSphs.
Each panel corresponds to the Fermi-LAT upper bounds derived for a different DM density profile in an optimised ROI. Colours represent 
different annihilation channels in the Galactic halo, $b\bar{b}$ final states are displayed in grey, $c\bar{c}$ in green, $H^0_1H^0_1$ in blue, $H^0_1H^0_2$ in dark blue, $A^0_1A^0_1$ in cyan, $gg$ in yellow, 
$\tau^+\tau^-$ in dark red, $W^+W^-$ in violet and $NN$ in dark green. } 
\label{fig:sigvgg}
\end{figure}

In Figure \ref{fig:sigvgg}, we show the thermally averaged annihilation cross section of RH sneutrinos in the Galactic halo for the di-photon final state, 
$\sigmavgg$, as a function of the RH sneutrino mass, for the solutions that satisfy the different experimental constraints mentioned in Section\,\ref{sec:neutscan}.
Black lines represent the Fermi-LAT bounds at 68\% and 95\%CL  (black dotted and thin black lines respectively). 
Each panel corresponds to a different assumption of the DM density profile, the top left panel corresponds to a NFW, top right to an Einasto,
bottom left panel to a NFW with adiabatic contraction (NFW$_c$), and bottom right to an isothermal density profile. 
Since the $J$-factor for each halo is different the bounds also vary significantly. 
Different colours in Fig.~\ref{fig:sigvgg}, represent different dominant RH sneutrino annihilation final states in the Galactic halo, although we stress that the gamma-ray flux has been computed for each point using the contributions for all channels. Thus,
$b\bar{b}$ channel is represented in grey, $c\bar{c}$ in green, $H^0_1H^0_1$ in blue, $H^0_1H^0_2$ in dark blue, $A^0_1A^0_1$ in cyan, $gg$ in violet, 
$\tau^+\tau^-$ in red, $W^+W^-$ in yellow and $NN$ (RH neutrinos) in dark green. 
For clarity, the predictions for each of the channels using an Einasto profile are shown separately in Figure \ref{fig:sigvgg_annchannels} of Appendix\,\ref{sec:suppl_rhn}.

As we can observe in Figure \ref{fig:sigvgg}, there is a substantial number of points which present an enhanced annihilation into monochromatic photons with respect to the typical value of $10^{-31}$~cm$^3$\,s$^{-1}$, some of them even exceeding the Fermi-LAT bounds. 
This is due to the enhancement of the contribution from the loop diagrams of Figure \ref{fig:diphotonann} under several circumstances. 
On the one hand, these terms are intensified near a resonance of the RH sneutrino with a CP-even Higgs due to the Breit-Wigner effect. This was used in Ref.\,\cite{Chatterjee:2014bva} in order to account for a gamma-ray line at 130~GeV from RH sneutrino annihilation, and in our more general scan these resonances occur throughout the whole mass range. In our scan, resonances occur with the lightest and second lightest CP-even Higgs bosons.
On top of this, we have also observed a threshold enhancement \cite{Jackson:2013pjq} when the particles involved in the loops become on-shell. This takes place mainly for loops with $W$ bosons (when $\snmassr\approx m_W$) and charginos (when $\snmassr\approx\mcha1\approx\mu$).

All these effects are more prominent in the RH sneutrino than in the neutralino case, mainly due to the larger flexibility in the Higgs sector,  which is mainly unaffected by variations in the mass and couplings of the RH sneutrino (other than through the invisible decay width). As a result the predictions for $\sigmavgg$ can be larger and a fraction of points in the mass range between 30 GeV and 120 GeV, approximately, exceed the Fermi-LAT bound from gamma ray lines search.

Moreover, the RH sneutrino annihilation into either a pair of very light pseudoscalar or a pair of scalar Higgs bosons are dominant annihilation channels when they are kinematically allowed \cite{Cerdeno:2011qv,Cerdeno:2014cda}. These Higgses can subsequently decay in-flight into a pair of photons, as already mentioned in Ref.\,\cite{Cerdeno:2015ega}. In our scan the annihilation cross section into $\higgsl\higgsl$ and $\phiggsl\phiggsl$ can be rather large, thereby entailing a sizable contribution to box-shaped features.
As we explained  in Ref.\,\cite{Cerdeno:2015ega}, in order to determine whether these features are observable in Fermi-LAT data, we approximate their contribution as a continuum of lines extending from the minimum to the maximum box energy.  For this reason, excluded points do not necessarily correspond to points with a $\sigmavgg$ above the Fermi-LAT bound. 
In fact, we have observed that the number of examples excluded by box-shaped features is larger than those excluded by  $\sigmavgg$. 
This constitutes another important difference with the neutralino case, for which these contributions were more difficult to enhance.

Finally, we have also found points with a simultaneous enhancement of the  $\higgs{i}gg$ and $\higgsi\gamma\gamma$ effective couplings when a singlet-like scalar Higgs with $|S^{2}_{H^0_i}/S^{1}_{H^0_i}|\gtrsim5\tanb$ is present. In principle, these points might be tested in LHC searches for new light Higgs bosons decaying into two photons, since Higgs production through gluon fusion is enhanced with respect to the SM expectation.

\begin{figure}[t!]
	\begin{center}  
	\epsfig{file=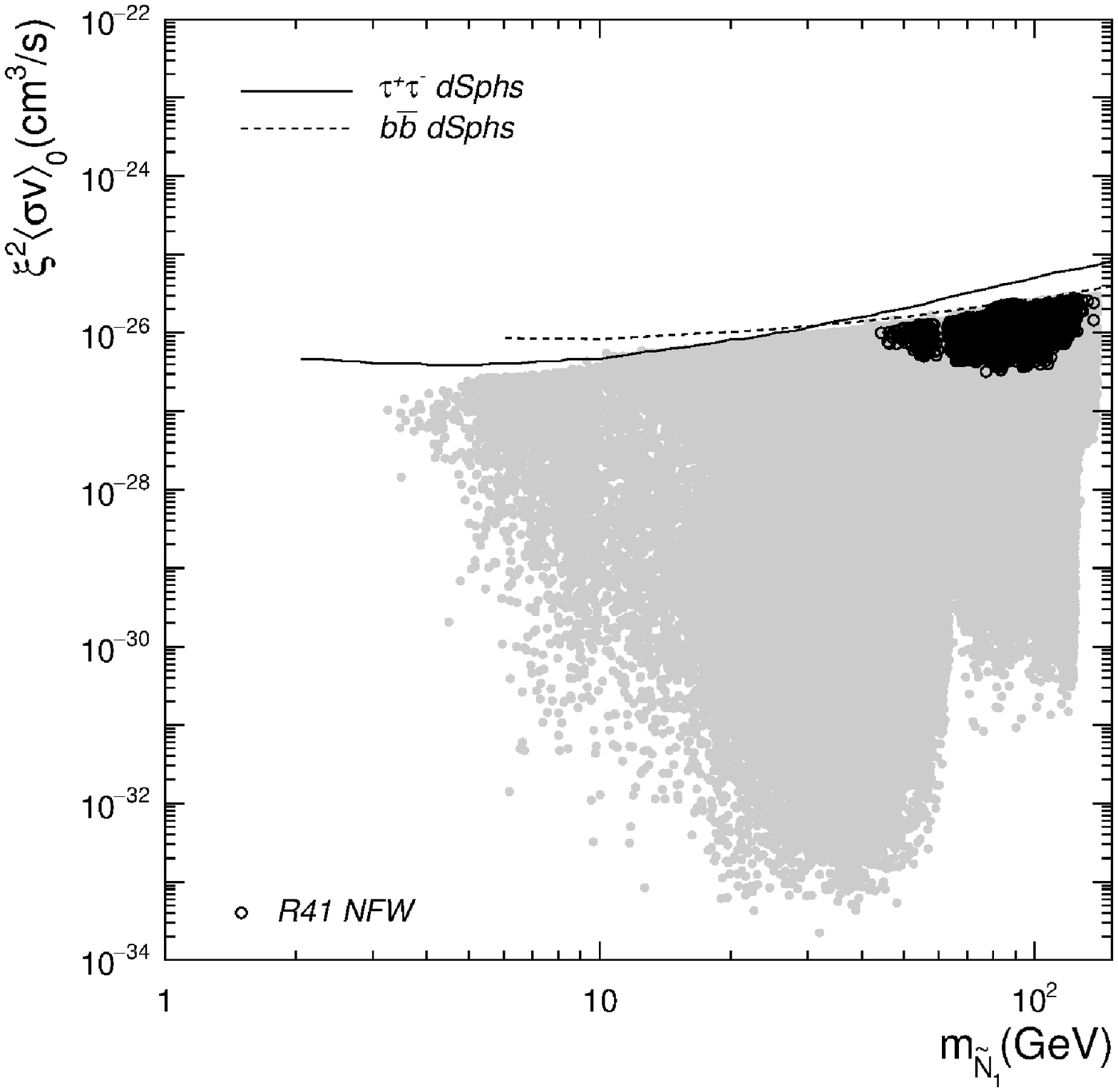,width=7.5cm}
	\epsfig{file=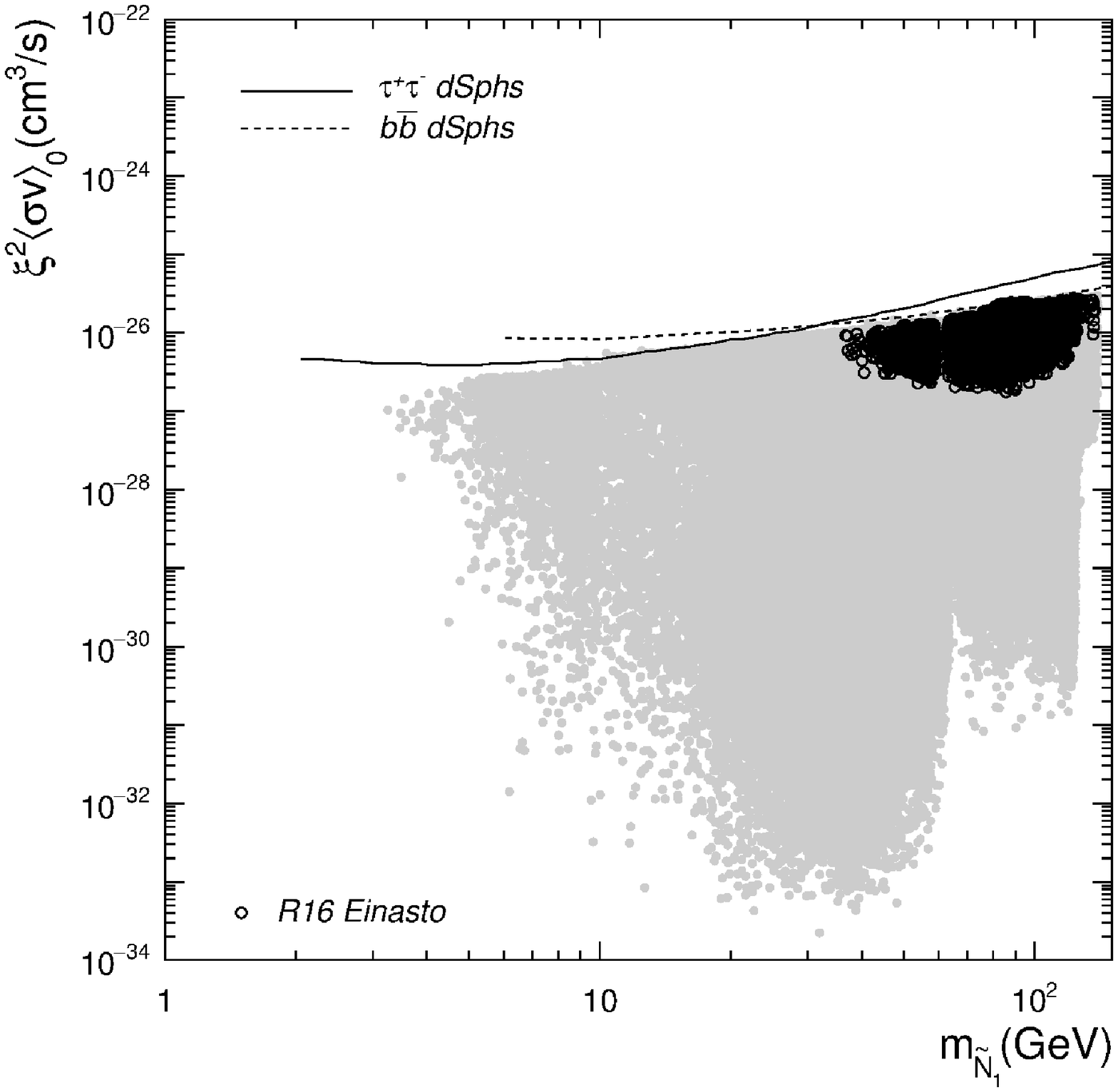,width=7.5cm}\\
	\epsfig{file=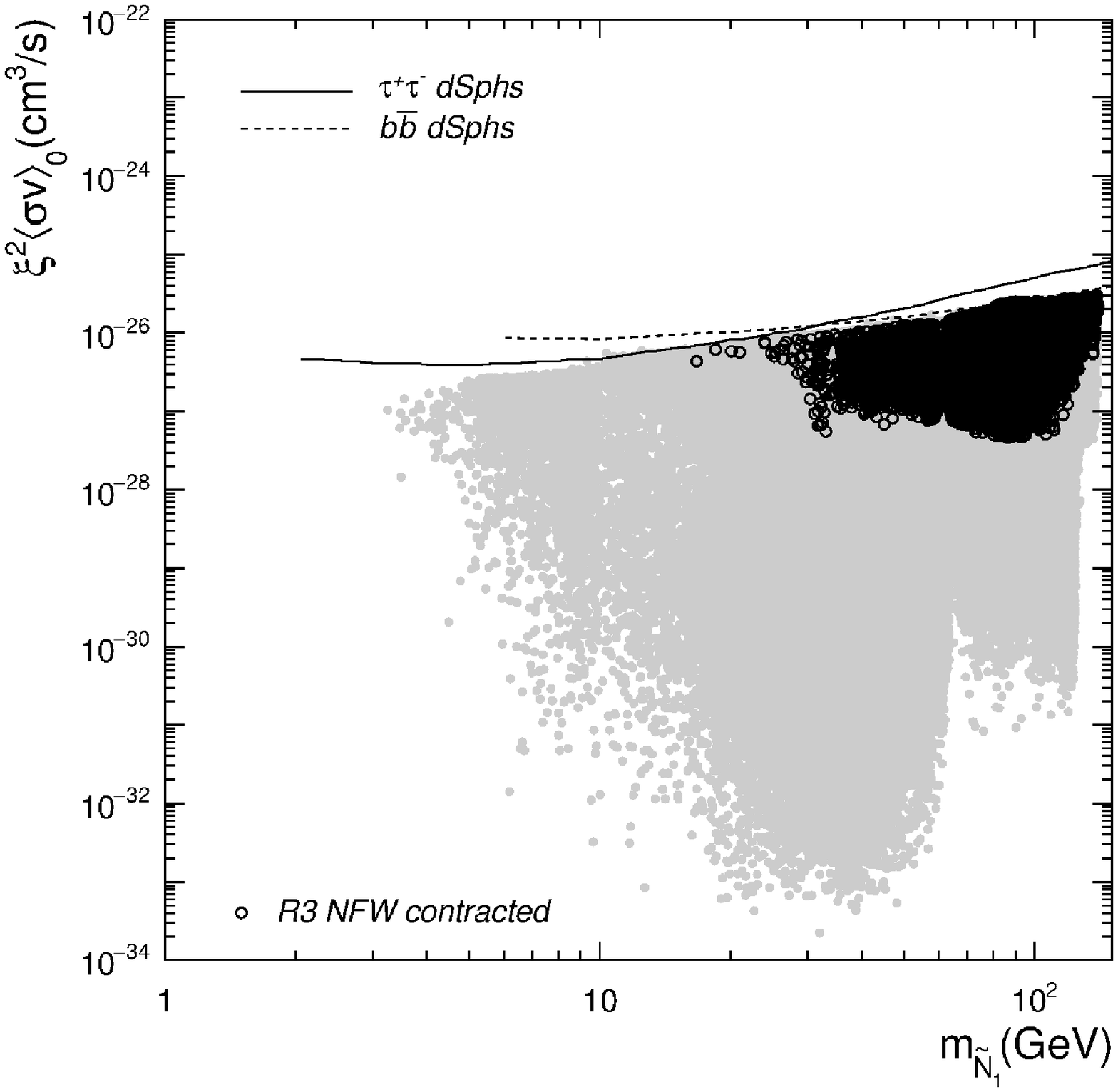,width=7.5cm}
	\epsfig{file=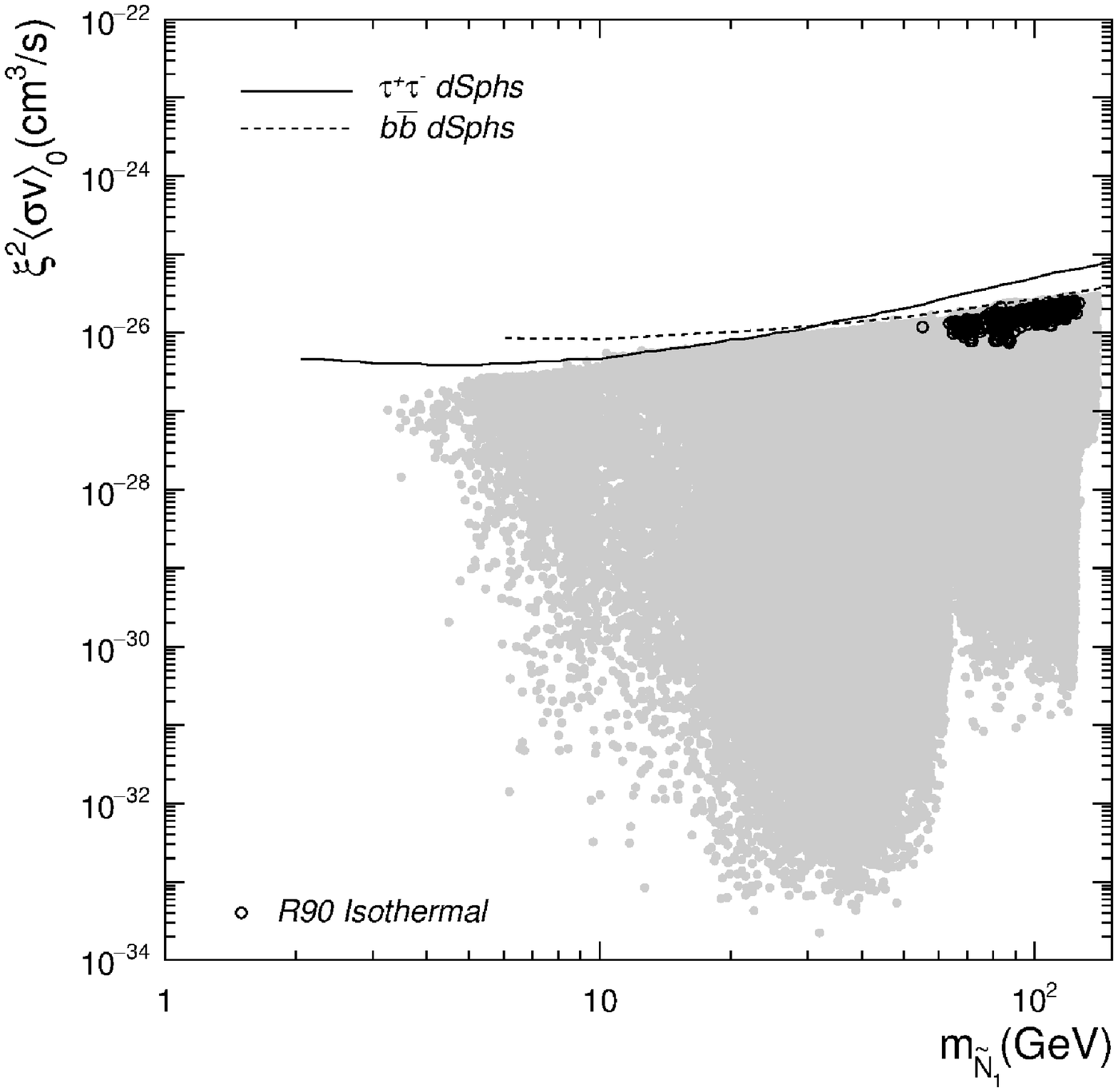,width=7.5cm}
	\end{center}
\caption{\small Thermally averaged RH sneutrino annihilation cross section in the Galactic halo as a function of the RH sneutrino mass. 
All the experimental constraints, including bounds from direct detection experiments and Fermi-LAT dSph data have been considered. 
The solid and dashed lines correspond to the upper bounds on $\langle\sigma v\rangle_0$ derived from an analysis of dSph galaxies for pure $\tau^-\tau^+$ and $b\bar b$ channels, respectively. 
Black circles correspond to points whose $\xi^2\langle\sigma v\rangle_{\gamma\gamma}$ for lines and/or box-shaped features exceeds the Fermi-LAT bounds for each DM density profile considered.}
\label{fig:sigv}
\end{figure}

Let us now compare these results with the Fermi-LAT bounds from dSphs. Figure\,\ref{fig:sigv} shows the theoretical predictions for the annihilation cross section of RH sneutrinos in the Galactic halo as a function of the RH sneutrino mass for the same choices of DM density profiles.
All the experimental constraints, including the one from Fermi-LAT dSph data have been considered. 
As a reference, the Fermi-LAT upper bounds from dSphs assuming pure annihilation into $\tau^+\tau^-$ or $b\bar b$ (solid and dotted lines respectively) are also displayed. 
Black dots correspond to the points of the parameter space for which the monochromatic gamma-ray emission and/or the contribution from box-shaped features exceed the Fermi-LAT bounds, whereas grey dots correspond to the rest of the points without distinguishing annihilation channels.
As we can observe, contrary to what happened in the neutralino case, the bound on $\sigmavgg$ can be more constraining that the bound on the continuum gamma-ray emission for a significant region of the parameter space. The amount of excluded points depend slightly on the choice for DM halo (NFWc being obviously more constraining).

\begin{figure}[t!]
	\begin{center}  
     \epsfig{file=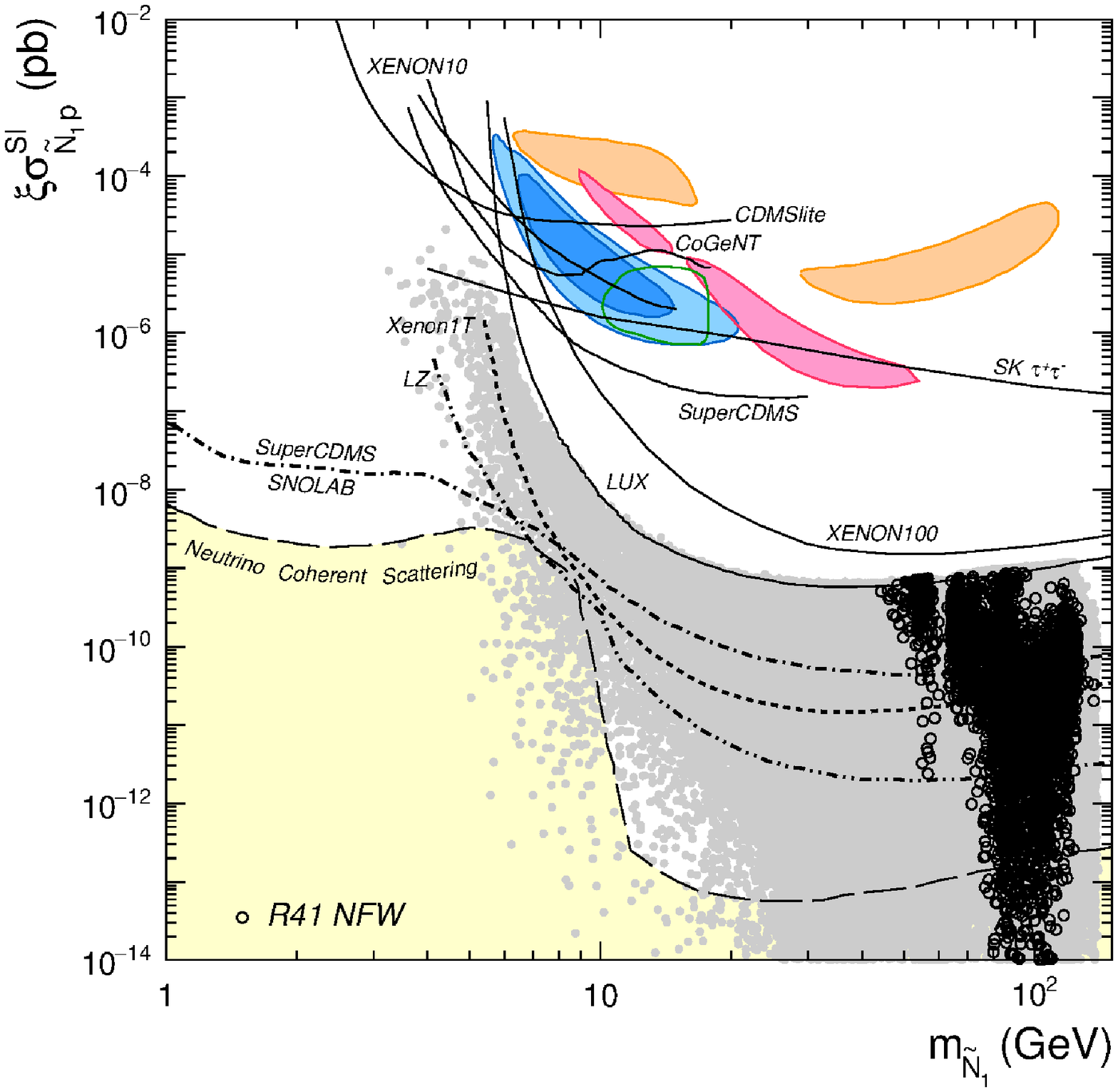,width=7.5cm}
	 \epsfig{file=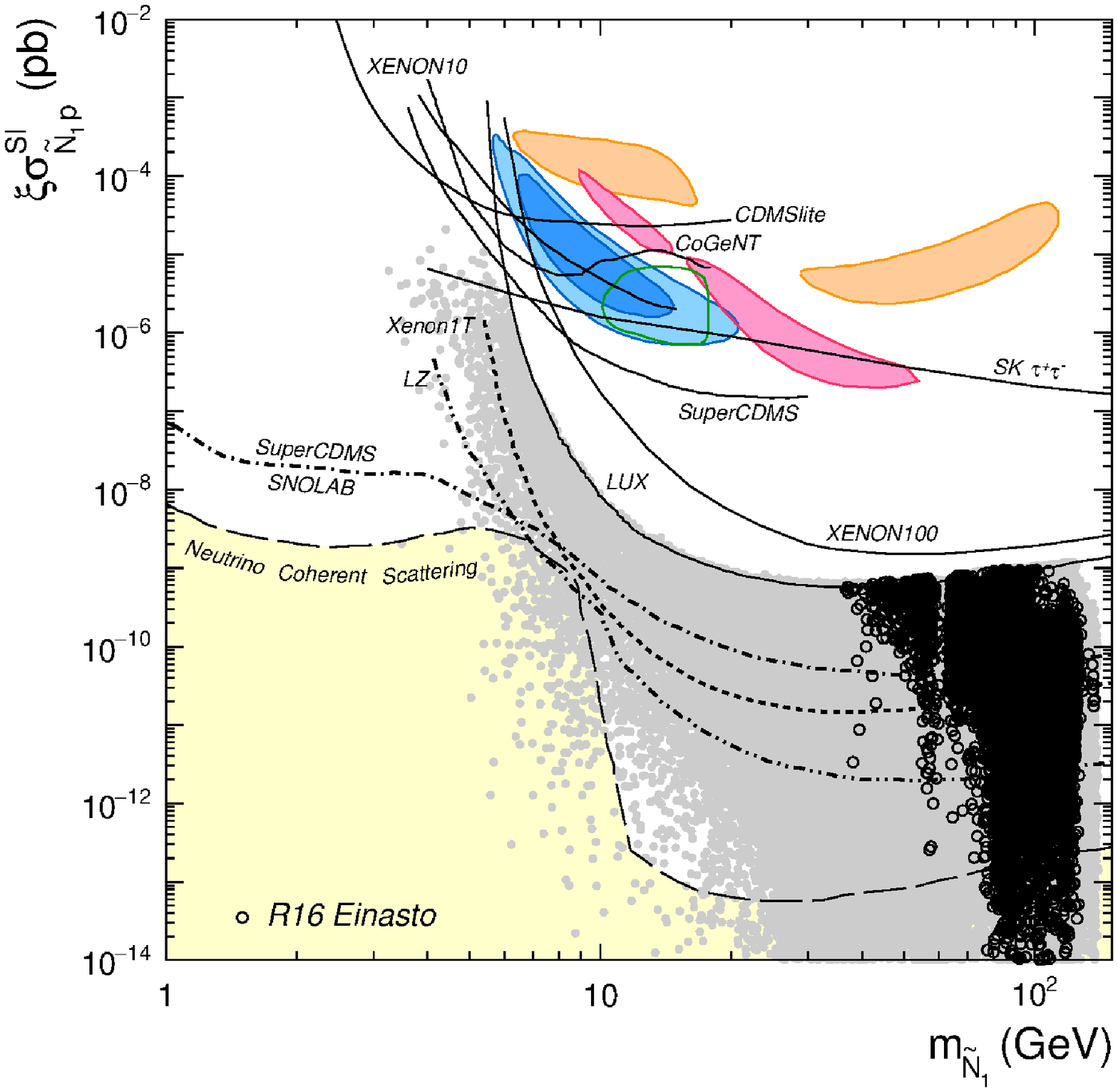,width=7.5cm}
     \epsfig{file=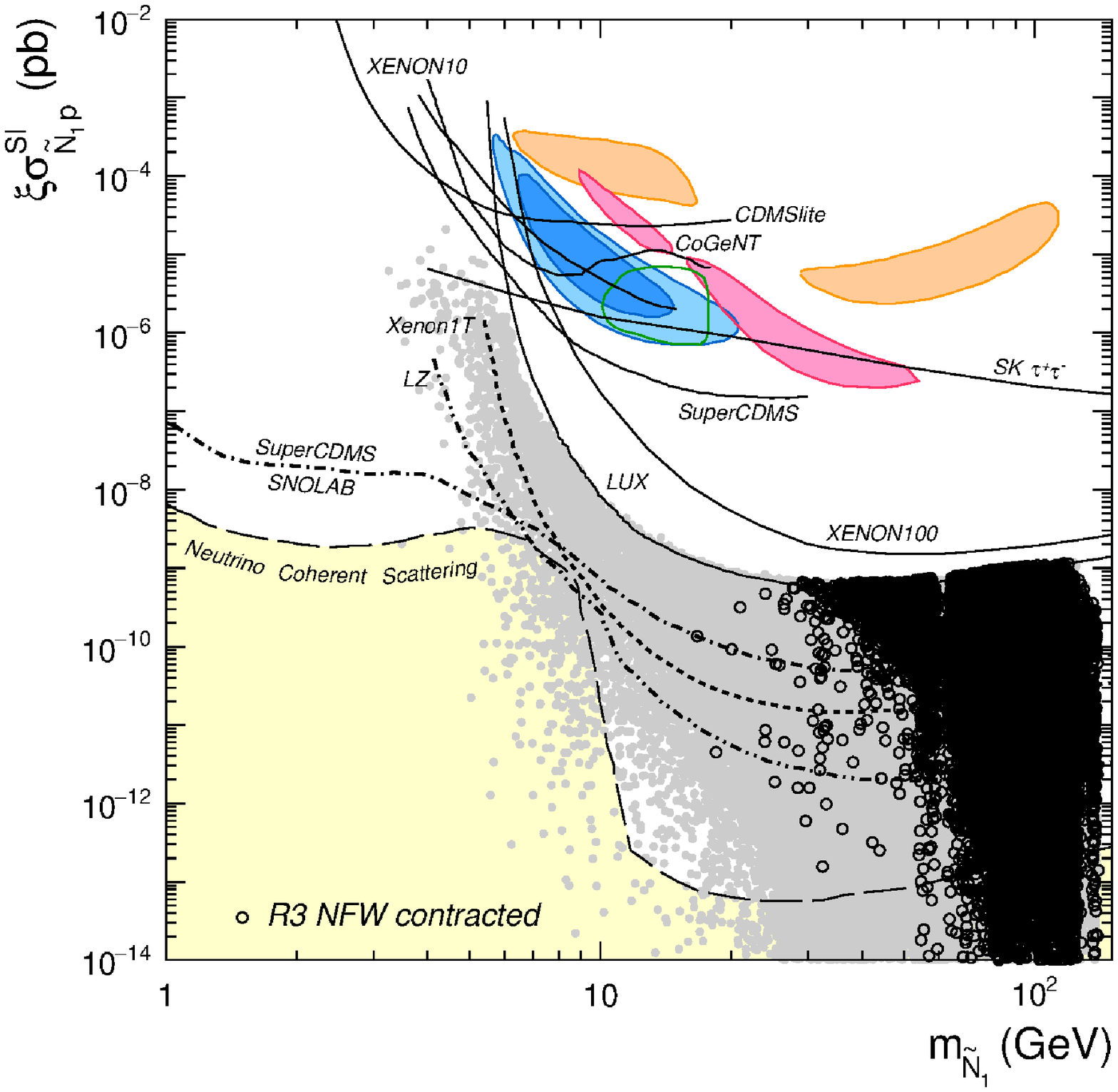,width=7.5cm}
     \epsfig{file=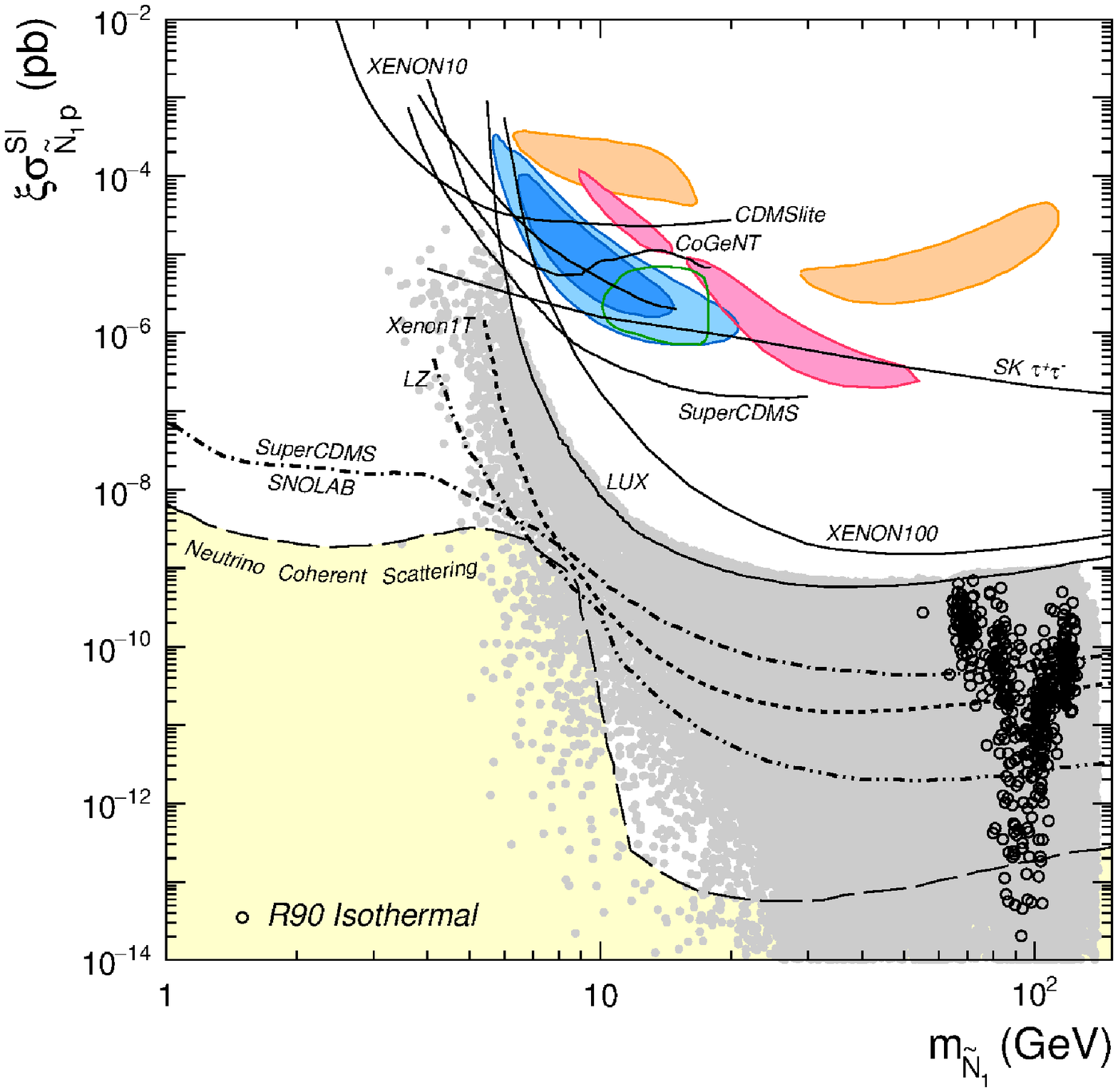,width=7.5cm}
	\end{center}
\caption{\small Spin independent sneutrino-proton cross section as a function of the sneutrino mass.
All the experimental constraints, including bounds from direct detection experiments and Fermi-LAT data on dSphs have been considered. 
Solid lines represent the current  experimental upper bounds from direct detection experiments, whereas dotted lines are the projected sensitivities of next-generation detectors. 
Both of them correspond to a rescaled SHM.
The dashed line corresponds to an approximate band where neutrino coherent scattering with nuclei will begin to limit the sensitivity of direct detection experiments. 
Closed contours represent the areas compatible with the observed excesses in  DAMA/LIBRA (orange), CRESST (red), CDMS II (blue), and CoGeNT (green). 
Black circles correspond to points whose $\xi^2\langle\sigma v\rangle_{\gamma\gamma}$ for lines and/or box-shaped features exceeds the Fermi-LAT bounds for each DM density profile considered.
}
\label{fig:si}
\end{figure}

Lastly, in Figure \ref{fig:si} we have plotted the theoretical predictions for the spin independent RH sneutrino-proton cross section and compared it with current and future direct detection bounds (the RH sneutrino, being a scalar field, has no contribution from spin-dependent interactions).
Once more, the points ruled out by the Fermi-LAT search for spectral features are shown in black and each panel represent a different halo profile like in the previous figure. 
It should be stressed that since the RH sneutrino scattering off quarks is mediated by a Higgs boson in the $t$-channel at tree level, in general we expect the RH sneutrino couplings to protons, $f_p$, and neutrons, $f_n$, to be very similar, resulting in $\sigma^{SI}_{\snr_1p}\approx\sigma^{SI}_{\snr_1n}$. Notice however that the different composition of protons and neutrons in valence quarks, $u$ and $d$, favours the occurrence of accidental cancellations in either coupling when the matrix elements of the scalar Higgs boson have different signs (an effect that has been already described for the neutralino \cite{Ellis:2000ds}). Due to the large flexibility of the Higgs sector in this construction, such ``blind spots" are frequent in our analysis. The ratio $f_n/f_p$ can be sizable (or very small) in such points \cite{Cheung:2012qy} (although the error induced by nuclear uncertainties \cite{Crivellin:2015bva} is also large). For consistency, as explained above, we are placing direct detection constraints on the total number of events, thereby including the contributions from protons and neutrons in a consistent way.
As we did in the case of the neutralino, the upper bounds for each halo have been computed using the procedure detailed in Ref.\,\cite{Marcos:2015dza}.
Remarkably, Fermi-LAT bounds on gamma-ray spectral features can rule out many solutions which escape current direct detection bounds and are beyond the reach of future experiments.

To summarise, contrary to what we have observed in the neutralino case, the contribution of RH sneutrino annihilation to spectral features in the gamma ray spectrum can exceed the Fermi-LAT constraints for a wide region of the parameter space. 
On the one hand, the Breit-Wigner effect near the resonances with CP even Higgses can lead to a sizable annihilation cross section into two photons. More importantly, the occurrence of CP even and CP odd Higgs final states, and their subsequent decay in-flight gives rise to observable box-shaped features.
Consequently, the bounds on spectral features can be more constraining than those on dSphs and direct DM detection.


\section{Conclusions}
\label{sec:conclusions}

In this article, we have computed the annihilation cross section in the Galactic halo of
two supersymmetric DM candidates: the neutralino in the NMSSM, and the RH sneutrino in an extended version of the NMSSM. 
For both models, we have studied the resulting gamma-ray spectrum, paying especial attention to the contributions from lines (monochromatic gamma-rays) and box-shaped spectral features.
We have compared the results with the upper bounds placed by the Fermi-LAT satellite, using the recent Pass 8 data set. 
These results have been contrasted with the ordinary search for a continuous emission of gamma-rays from dSphs, as well as with direct detection experiments.

To this aim, a series of scans over the parameter space of both models were conducted in the region with DM masses below 200~GeV. All the recent experimental constraints have been applied. We have implemented the LHC bounds on the masses of supersymmetric particles, Higgs mass and couplings, and low-energy observables such as the branching ratios of rare processes.
In addition, we have incorporated direct detection constraints on the elastic DM-nucleus cross section from the SuperCDMS and LUX experiments, computed according to the different DM density profiles of the Fermi-LAT ROIs optimised for gamma-ray line searches.
Finally, we have considered the Fermi-LAT bounds on dSphs. 

In the case of the lightest neutralino in the NMSSM, the majority of the viable points of the scanned parameter space occur in the vicinity of the resonances with either a light pseudoscalar, a scalar Higgs boson or a $Z$ boson.  
In agreement with previous analyses, we have found that the resulting annihilation cross section into two photons is generally very small, beyond the sensitivity of the Fermi-LAT. Only a reduced number of points would be observable thanks to a Breit-Wigner enhancement, which is more effective close to the resonance with the $Z$ boson, when $m_{\neutl}\approx45$~GeV, due to its larger decay width.
Similarly, the contribution from box-shaped features in the spectrum is beyond the reach of Fermi-LAT.
Even though the predictions of indirect detection through the contribution to the continuum gamma-ray flux are also below the current bounds from dwarf spheroidal galaxies, we have found that many of these points can be within the sensitivities of future direct detection experiments.
Additionally, we have provided theoretical predictions for the spin-independent and spin-dependent scattering cross section and compared them with the SuperCDMS and LUX results.

On the other hand, in the case of the RH sneutrino in an extended NMSSM, we have found that the search for spectral features can be much more constraining. 
Due to the flexibility in the Higgs sector, the RH sneutrino resonant annihilation through a scalar Higgs is much easier to achieve than in the case of the neutralino, hence the Breit-Wigner effect near the resonance can increase the annihilation cross section into a diphoton final state for a wide range of masses. 
On top of this, RH sneutrino annihilation into light pseudoscalar and scalar Higgses, and the subsequent decay in flight of these Higgses into pairs of photons give rise to box-shaped features in the spectrum that can be within the reach of Fermi-LAT. 
As a consequence, the search for features in the gamma-ray spectrum by Fermi-LAT can be more constraining than Fermi-LAT bounds from dSphs for sneutrino masses in the approximate range between 30 and 120 GeV.

Many of the points that can be probed through the search for spectral features might escape direct detection, thus our findings reinforce the importance of a complementary strategy for DM searches. Moreover, the very different predictions for neutralinos and RH sneutrinos demonstrate that the observation of spectral features in the gamma-ray spectrum might not just be a smoking gun for DM annihilation, but also could help to discriminate among different WIMP candidates.

\clearpage

\appendix

\noindent{\Large\bf Appendix}


\section{Matrix elements for RH sneutrino annihilation into two photons}
\label{sec:oneloopcalc}

We include here the matrix elements corresponding to the RH sneutrino annihilation into two photons. The one-loop diagrams have been shown in Figure \ref{fig:diphotonann}, and we make use of the RH sneutrino vertices that can be found in Ref.~\cite{Cerdeno:2009dv}.
Hereafter, we will express our calculations in terms of the function $f(\tau)$ defined in Ref.~\cite{Gunion:1989we},
\begin{equation}
f(\tau)=\begin{cases}
         \arcsin^2(\tau^{-1/2}) & \mathrm{for} \quad \tau \geq 1 \\
         -\frac14 \left[\log\frac{1+\sqrt{1-\tau}}{1-\sqrt{1-\tau}}-i\pi \right]^2 &  \mathrm{for} \quad  \tau < 1 
         \end{cases},
\nonumber
\label{funcf}
\end{equation}
and for a given particle $X$, we define $\tau_X\equiv\taufrac{X}$. In the following, we consider $\tau_X \geq 1$.

\begin{itemize}
 
\item Annihilation mediated by a $\chiggspm$ loop (corresponding to the first row of 
Fig.~\ref{fig:diphotonann})
\begin{equation}
|\overline{\CM}_{\gamma\gamma}^{\chiggspm loop}|^2=\frac{\alpem^2}{2\pi^2}\cccsnsn^2
\left[1-\tauh f(\tauh)\right]^2 .
\nonumber
\end{equation}

\item Annihilation mediated by a $\higgsi$

\paragraph{- Top quark loop.}
\begin{eqnarray}
|\overline{\CM}_{\gamma\gamma}^t|^2 &=&\frac{2N_c^2Q_t^4\alpem^2g^2m_t^4}{\pi^2\mw^2\sinbsq}
\left[1+\left(1-\taut\right)f(\taut)\right]^2\times \nonumber \\
&&\left|\sum_{i,j=1}^3\frac{\hcompiu\chisnsn\hcompju\chjsnsn}{(\props{i})(\cprops{j})}\right| ,
\nonumber
\end{eqnarray}
where $Q_t$ is the top charge and $N_c$ is the flavour number.

\paragraph{- $\mathbf{W^\pm}$ loop.} 
\begin{eqnarray}
|\overline{\CM}_{\gamma\gamma}^W|^2 &=&\frac{\alpem^2g^2s^2}{32\pi^2\mw^2}[2+3\tauw+3(2\tauw-\tauw^2)f(\tauw)]^2
\times\nonumber\\
&&\left|\sum_{i,j=1}^3\frac{\ch{i}\chisnsn\ch{j}\chjsnsn}{(\props{i})(\cprops{j})}\right| .
\nonumber
\end{eqnarray}

\paragraph{- $\mathbf{\chiggspm}$ loop.} 
\begin{eqnarray}
|\overline{\CM}_{\gamma\gamma}^\chiggspm|^2=\frac{\alpem^2}{2\pi^2}\left[1-\tauh f(\tauh)\right]^2
\left|\sum_{i,j=1}^3\frac{\chisnsn\chcc{i}\chjsnsn\chcc{j}}{(\props{i})(\cprops{j})}\right| .
\nonumber
\end{eqnarray}

\paragraph{- Sfermion loop.} The leading contributions are expected to come from stops, sbottoms and staus.
\begin{eqnarray}
|\overline{\CM}_{\gamma\gamma}^{\st}|^2 &=&\frac{N_c^2Q_t^4\alpem^2}{2\pi^2} 
\sum_{a,b=L,R}
[1-\taufa f(\taufa)][1-\taufb f(\taufb)]\times\nonumber\\
&&\left|\sum_{i,j=1}^3\frac{\chisnsn\chstst{i}{a}\chjsnsn\chstst{j}{b}}{(\props{i})(\cprops{j})}\right| ,
\nonumber
\end{eqnarray}
where 
the $\chstst{i}{a}$ coupling can be found in Ref.~\cite{Franke:1995tc}.

\paragraph{- Chargino loop.} 
The squared matrix element can be written as
\begin{eqnarray}
|\overline{\CM}_{\gamma\gamma}^{\chg}|^2 &=&\frac{8\alpem^2}{\pi^2}
\sum_{a,b=1}^2
\left[1+\left(1-\tauxa\right)f(\tauxa)\right]
\left[1+\left(1-\tauxb\right)f(\tauxb)\right]
\times\nonumber\\
&&\left|\sum_{i,j=1}^3\frac{\chisnsn\ccchch{i}{a}{a}\mcha{a}\chjsnsn\ccchch{j}{b}{b}\mcha{b}}{(\props{i})(\cprops{j})}\right| ,
\nonumber
\end{eqnarray}
where $i\ccchch{i}{a}{b} \equiv-i(\Qab{i}{a}{b}^*P_L+\Qab{i}{b}{a}P_R)$. The definition of the $Q$ matrix can be found in Ref.~\cite{Franke:1995tc}.

\end{itemize}


\section{Supplementary plots for RH sneutrino DM}
\label{sec:suppl_rhn}

\begin{figure}[t!]
	\begin{center}  
	\epsfig{file=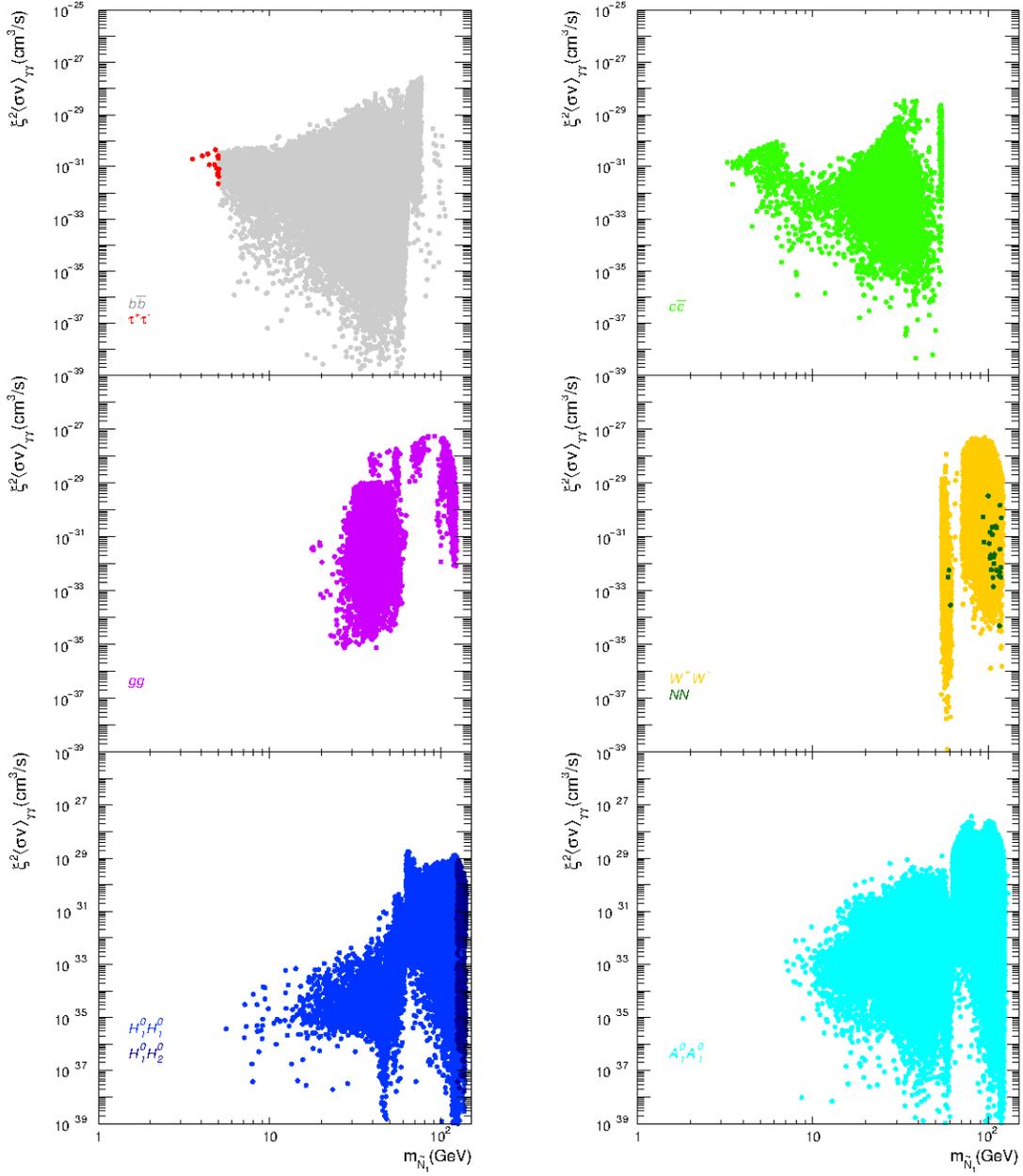,width=15cm}
	\end{center}
\caption{\small Thermally averaged RH sneutrino annihilation cross section into two photons in the Galactic halo versus the sneutrino mass. 
All the points fulfil the experimental constraints, including bounds from direct detection experiments and Fermi-LAT data on dSphs, and have a relic abundance $0.001<\Omega_{\tilde{N}_1}h^2<0.13$.
Each colour represents different dominant annihilation channels.} 
\label{fig:sigvgg_annchannels}
\end{figure}

For clarity, we include in this appendix some complementary plots that help to understand how the different annihilation channels are distributed in the parameter space. In particular, Figure \ref{fig:sigvgg_annchannels} displays the thermally averaged RH sneutrino annihilation cross section in the DM halo, as in Figure \ref{fig:sigvgg}, but where the points with different dominant annihilation channels are presented separately (thus avoiding overlapping). For example, this figure shows that $b\bar b$ is widely extended for RH sneutrino masses below 60~GeV, and that $\higgsl\higgsl$ and $\phiggsl\phiggsl$ are also easy to obtain.

\begin{figure}[t!]
	\begin{center}  
	\epsfig{file=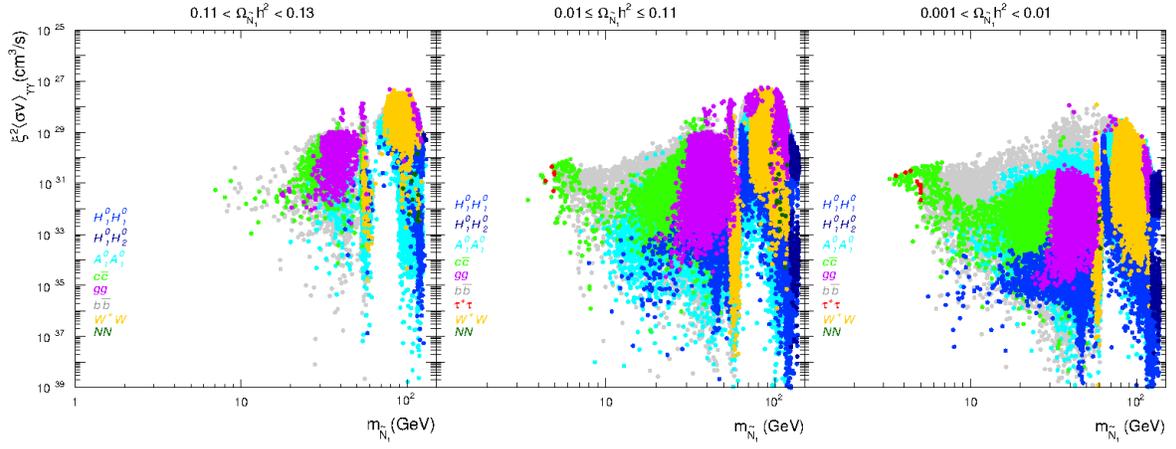,width=16cm}
	\end{center}
\caption{\small Thermally averaged RH sneutrino annihilation cross section into two photons in the Galactic halo versus the sneutrino mass. 
From left to right, the three panels show different ranges in the RH sneutrino relic abundance, the leftmost one being in agreement with Planck results.
All the experimental constraints are included, together with the bounds from direct detection experiments and Fermi-LAT data on dSphs. The colour code is as in Figure \ref{fig:sigvgg}. 
} 
\label{fig:sigvgg_relic}
\end{figure}

Finally, in Figure \ref{fig:sigvgg_relic} we separate our results in three ranges in the RH sneutrino abundance. If we restrict ourselves to values of $\Omega_{\tilde{N}_1}h^2$ compatible with the total DM abundance, many of the points below $\snmassr\sim10$~GeV disappear. Otherwise there is no appreciable change. Notice that underabundant DM scenarios imply larger values for annihilation cross section. However, since we consider a suppression factor $\xi^2$, the resulting gamma-ray flux ends up being slightly suppressed, as we can observe comparing the central and right-hand panels.

\clearpage

\noindent{\bf \large Acknowledgements}

D.G.C acknowledges support from the STFC and the partial support of the Centro de Excelencia Severo Ochoa Program through the IFT UAM/CSIC.
S.R. is supported by Campus of Excellence UAM+CSIC.
We thank the support of the Consolider-Ingenio 2010 program under grant MULTIDARK CSD2009-00064, the Spanish MICINN under Grants No. FPA2012-34694 and FPA2013-44773-P, the Spanish MINECO Centro de Excelencia Severo Ochoa Program under Grant No. SEV-2012-0249,
and the European Union under the ERC Advanced Grant SPLE under contract ERC-2012-ADG-20120216-320421.

\bibliography{2015-cpr-v18}

\end{document}